\documentclass[11pt]{article}
\pdfoutput=1
\usepackage{jheppub}
\usepackage{amsmath,amssymb,braket,bbold,graphicx,color,ulem}
\usepackage{empheq}
\usepackage{todonotes} 
\usepackage{tikz} 
\usetikzlibrary{trees,decorations.pathmorphing,decorations.markings} 
\usepackage{booktabs} 

\newcommand{\be}{\begin{equation}}
\newcommand{\ee}{\end{equation}}
\newcommand{\bea}{\begin{eqnarray}}
\newcommand{\eea}{\end{eqnarray}}
\newcommand{\beqn}{\begin{eqnarray}}
\newcommand{\eeqn}{\end{eqnarray}}
\newcommand{\nn}{\nonumber}

\def\la{\langle}
\def\ra{\rangle}

\newcommand{\tabra}[1]{\renewcommand{\arraystretch}{#1}}

\newcommand{\td}{\mathrm{d}}

\newcommand{\p}{\partial}
\newcommand{\ph}{\phantom}

\newcommand{\gmn}{g_{\mu\nu}}
\newcommand{\fmn}{f_{\mu\nu}}
\newcommand{\bgmn}{\bar g_{\mu\nu}}
\newcommand{\bfmn}{\bar f_{\mu\nu}}
\newcommand{\hmn}{h_{\mu\nu}}
\newcommand{\lmn}{\ell_{\mu\nu}}
\newcommand{\dG}{\delta G}
\newcommand{\dM}{\delta M}

\newcommand{\dd}{\mathrm{d}}
\newcommand{\mpl}{m_\mathrm{Pl}}
\newcommand{\mfp}{m_{\rm FP}}
\newcommand{\Lag}{\mathcal{L}}

\newcommand{\GeV}{\mathrm{GeV}}
\newcommand{\TeV}{\mathrm{TeV}}

\tikzset{
    graviton/.style={decorate, decoration={coil,amplitude=4pt, segment length=5pt}, semithick, double, draw=red},
    Mgraviton/.style={decorate, decoration={snake,amplitude=5pt}, semithick, double, draw=magenta},
    sm/.style={draw=blue,double,thick},
}

\title{Heavy spin-2 Dark Matter}
\author[1,2]{Eugeny~Babichev,} 
\author[3,4]{Luca~Marzola,}
\author[3,4]{Martti~Raidal,}
\author[5]{Angnis~Schmidt-May,}
\author[3]{Federico~Urban,}
\author[3]{Hardi~Veerm\"ae,}
\author[2
]{Mikael~von~Strauss}

\affiliation[1]{Laboratoire de Physique Th\'eorique, CNRS, Univ.~Paris-Sud, Universit\'e Paris-Saclay,\\ 91405 Orsay, France}
\affiliation[2]{UPMC-CNRS, UMR7095,
 Institut d'Astrophysique de Paris, GReCO,\\
 98bis boulevard Arago, F-75014 Paris, France.}
 \affiliation[3]{National Institute of Chemical Physics and Biophysics, R\"avala 10, 10143 Tallinn, Estonia.}
\affiliation[4]{Laboratory of Theoretical Physics, Institute of Physics, University of Tartu,\\
 Ravila 14c, 50411 Tartu, Estonia.}
\affiliation[5]{Institut f\"ur Theoretische Physik, Eidgen\"ossische Technische Hochschule Z\"urich,\\
Wolfgang-Pauli-Strasse 27, 8093 Z\"urich, Switzerland}

\emailAdd{eugeny.babichev@th.u-psud.fr}         
\emailAdd{luca.marzola@ut.ee}
\emailAdd{martti.raidal@cern.ch}
\emailAdd{angniss@phys.ethz.ch}
\emailAdd{federico.urban@kbfi.ee}
\emailAdd{hardi.veermae@ut.ee}
\emailAdd{strauss@iap.fr}

\abstract{
We provide further details on a recent proposal addressing the nature of the dark sectors in cosmology and demonstrate that all current observations related to Dark Matter can be explained by the presence of a heavy spin-2 particle. Massive spin-2 fields and their gravitational interactions are uniquely described by ghost-free bimetric theory, which is a minimal and natural extension of General Relativity. In this setup, the largeness of the physical Planck mass is naturally related to extremely weak couplings of the heavy spin-2 field to baryonic matter and therefore explains the absence of signals in experiments dedicated to Dark Matter searches. It also ensures the phenomenological viability of our model as we confirm by comparing it with cosmological and local tests of gravity. At the same time, the spin-2 field possesses standard gravitational interactions and it decays universally into all Standard Model fields but not into massless gravitons. Matching the measured DM abundance together with the requirement of stability constrains the spin-2 mass to be in the 1 to 100 TeV range.
}


\begin{document}

\unitlength=1mm
\maketitle \flushbottom

\section{Introduction}\label{sec:intro}

Numerous cosmological and astrophysical observations have confirmed the presence of a Dark Matter (DM) component in our Universe. Until now this unknown type of matter has been seen only through its gravitational interactions, which resemble those of ordinary matter. Its effects are visible in the rotation curves and velocity dispersions of galaxies, gravitational lensing, matter distribution power spectra, structure formation, Baryon Acoustic Oscillations and angular power spectrum of the Cosmic Microwave Background~\cite{Agashe:2014kda}. 
The standard paradigm treats the unknown DM particle as a cold relic density which has been created through a model-dependent production mechanism in the early Universe. General Relativity (GR) as the theory for gravity (including a cosmological constant $\Lambda$ which accounts for the observed amount of Dark Energy) together with a particle physics model for cold Dark Matter (CDM) yield the concordance description of cosmology, the $\Lambda$CDM model. See \cite{Bull:2015stt} for a recent review of its status quo.

The most popular DM models moreover assume that the DM particle is weakly coupled to baryonic matter and hence might be produced in colliders, directly detected in dedicated experiments or indirectly observed through astro-particle signatures. From a theoretical perspective, many of these models lose some of their attractiveness because, typically, they are either not very well motivated from fundamental principles or they introduce a large number of unobserved additional fields (such as Supersymmetry). Unfortunately, on the experimental side, all attempts to produce or detect the DM particle have remained unsuccessful so far~\cite{Agashe:2014kda, Ackermann:2015zua, Ahnen:2016qkx, Ackermann:2015lka, Khachatryan:2014rra, Gaskins:2016cha}.

The absence of any signatures for DM apart from its gravitational effects motivates a shift of paradigm in the way to think about the nature of DM. Instead of augmenting the Standard Model (SM) by an additional field, we suggest that the DM particle may instead arise in a minimal extension of the gravitational sector, namely in the form of an additional massive spin-2 field. To us this seems to be a natural and well-motivated proposal, since there is no evidence supporting the fact that DM shares the quantum numbers of one of the SM particles and we only observe it through its gravitational interactions.

General Relativity can be treated as the unique theory of a single massless spin-2 particle, the graviton. We will colloquially refer to this point of view as the standard description of gravity. Since massless spin-2 fields cannot interact with each other~\cite{Boulanger:2000rq}, the most natural and minimal addition to gravity is that of a massive spin-2 field. Studying the effects of a massive spin-2 field in addition to standard gravity amounts to answering fundamental questions of field theory.
Remarkably, for several decades it was believed that no consistent theory for gravitating massive spin-2 fields can be formulated owing to the unavoidable presence of a fatal ghost instability~\cite{Boulware:1973my}. Only recently has the unique description which avoids the ghost been found~\cite{deRham:2010kj, Hassan:2011hr, Hassan:2011tf, Hassan:2011zd}. Since it involves an additional dynamical tensor field, which mixes with the gravitational metric, the corresponding theory has been titled ``bimetric theory of gravity". If fundamental massive spin-2 particles exist, they are described by this unique theory, which automatically leads to a modification of gravity. For the history and detailed reviews of theories for massive spin-2 fields we refer the reader to~\cite{Hinterbichler:2011tt, deRham:2014zqa, Schmidt-May:2015vnx}.

Following ideas outlined in \cite{Schmidt-May:2015vnx, Schmidt-May:2016hsx}, it has been proposed that the existence of a massive spin-2 particle can explain all the effects related to DM~\cite{Babichev:2016hir} (see also \cite{Aoki:2016zgp}). The present work is dedicated to providing details of and more insights into this novel proposal.

\paragraph{Summary of results.}

Being a modification of gravity, bimetric theory must satisfy constraints coming from Solar System tests and cosmology. We confirm that a large value for the spin-2 mass, together with a small value for the ``second Planck mass" of the metric that does not couple directly to matter, imply that the static spherically symmetric and cosmological solutions of bimetric theory always resemble those of GR. 
In this parameter region, where bimetric theory passes all observational tests of GR, the additional massive spin-2 field continues to gravitate but decouples from matter, automatically providing an ideal DM candidate.

We derive the conditions which ensure the validity of a perturbative treatment of bimetric theory for the interesting parameter regions and energy regimes. The structure of cubic and higher interactions for the spin-2 fields forbids a decay of the massive field into massless gravitons, resulting in a discriminating feature of the bimetric model. 

Requiring sufficient spin-2 DM to be produced in the early Universe and imposing constraints coming from its possible decay into SM fields, we obtain the allowed region in the bimetric parameter space: The spin-2 mass has to lie within the narrow region of $1~\TeV\lesssim\mfp\lesssim66~\TeV$ and the ratio of Planck masses must satisfy $10^{-11}\lesssim\alpha\lesssim10^{-15}$. 
This region overlaps with the one where classical solutions to the bimetric equations resemble GR to a very high precision and therefore the theory passes all observational tests. Moreover, our setup introduces no additional energy scale significantly higher than the weak scale and thus does not create any new hierarchy problems with respect to GR.

Our novel DM proposal naturally explains the absence of signals in (in)direct detection experiments and colliders. A prediction of the model is that any future experiments of this kind will continue to produce null-results.
In turn, a detection of a DM particle with mass below our predicted value would rule out our proposal of a heavy spin-2 as sole explanation for the observed DM. Alternative tests of our proposal could be based on its gravitational or its self-interacting nature.

\paragraph{Outline of the paper.}

Section~\ref{sec:BMreview} is dedicated to reviewing relevant details of bimetric theory. We provide its action, equations of motion and present the maximally symmetric solutions with the corresponding mass spectrum.
In section \ref{sec:BMtoGR}, we discuss the two parameter regions for which the classical solutions of the theory resemble those of GR.  The regime of parameters and energies where the bimetric action can be treated perturbatively is derived in the beginning of section~\ref{sec:heavymass}. Thereafter, we compute the cubic and quartic vertices in the spin-2 sector, verifying the absence of decay terms into massless gravitons. The phenomenology of spin-2 DM is explored in section~\ref{sec:pheno}. Finally, we discuss our results in section~\ref{sec:conclusions}. Additional supporting details can be found in the appendices.

\section{Details of bimetric theory of gravity}\label{sec:BMreview}

\subsection{Action and equations of motion}

In order to set down some notation and facilitate our later discussions we first provide some of the required basic details of the ghost-free bimetric theory. For further details and a recent review on the subject we refer to~\cite{Schmidt-May:2015vnx}. The theory is defined by the action \cite{Hassan:2011zd},
\begin{align}\label{bgaction}
S=m_g^2\int\td^4x\biggl[&\sqrt{|g|}R(g)+\alpha^2\sqrt{|f|}R(f)-2\alpha^2m_g^2\sqrt{|g|}\,V\left(S;\beta_n\right)\biggr]
+\int\td^4x\sqrt{|g|}\,\Lag_{\rm m}(g,\Phi)\,.
\end{align}
Here $m_g$ is a mass scale related to the reduced Planck mass via,
\be\label{defmpl}
\mpl^2=m_g^2(1+\alpha^2)\,,\qquad \mpl \approx 2.4\times10^{18}\,\GeV\,.
\ee
The dimensionless $\alpha^2$ (the ``ratio of Planck masses") measures the relative interaction strength of the two tensor fields.\footnote{As we will see later, $\alpha$ also quantifies the mixing between the propagating states relative to the interacting states. In the literature it is common to use mass parameters $m_g$ and $m_f$ in front of the kinetic terms as well as a mass parameter $m$ in front of the interaction terms. The scale of $m$ is redundant since it only parameterises the overall scale of the $\beta_n$ and can be chosen freely without loss of generality. In our notation, the relation to these additional mass parameters is given by $m_f=\alpha\,m_g$ and $m=\sqrt{m_gm_f}$.} In addition, the interaction potential $V(S;\beta_n)$ contains 5 dimensionless parameters $\beta_n$. Of these, $\beta_0$ and $\beta_4$ act as bare cosmological constants for $\gmn$ and $\fmn$, respectively, and therefore encode nonlinear self-interactions while the remaining parameters $\beta_1,\beta_2,\beta_3$ encode the nonlinear interactions between the two tensor fields.\footnote{However, all of the $\beta_n$ parameters contribute to the physical cosmological constant for constant curvature spacetimes, c.f.~Eq.~\eqref{eombimprop}.} The form of the interaction potential $V$ is constrained by demanding absence of the so called Boulware-Deser ghost~\cite{Boulware:1973my} and is given by~\cite{deRham:2010kj, Hassan:2011zd},
\be\label{Vdef}
V\left(S;\beta_n\right) = \sum_{n=0}^4\beta_n e_n(S)\,,
\ee
where the $e_n(S)$ are the elementary symmetric polynomials defined in terms of the eigenvalues of the matrix $S$. Explicitly they can be obtained via tracing the unit weight totally anti-symmetric products,
\be
e_n(S)=S^{\mu_1}_{~[\mu_1}\cdots S^{\mu_n}_{~\mu_n]}\,.
\ee
From the definitions it follows that $e_4(S)=\det(S)$ and that $e_n(S)=0$ for all $n>4$. The matrix argument $S$ which appears in the interaction potential is a square-root matrix defined through the relation,
\be\label{Sdef}
S^\rho_{~\sigma}S^\sigma_{~\nu}=g^{\rho\mu}\fmn\,.
\ee
For invertible $S$, the identity $e_n(S^{-1})=e_{4-n}(S)/e_4(S)$ can be used to show that,
\be
\sqrt{|g|}\,V(S;\beta_n)=\sqrt{|f|}\,V(S^{-1};\beta_{4-n})\,.
\ee
Apart from the matter interactions, the structure of the theory is therefore completely symmetric in terms of $\gmn$ and $\fmn$. In fact, in the absence of matter couplings the structure of the bimetric action is invariant under the following discrete interchanges,
\be\label{intsym}
\alpha^{-1}\gmn\leftrightarrow\alpha\fmn\,,\qquad
\alpha^{4-n}\beta_n\leftrightarrow \alpha^n\beta_{4-n}\,.
\ee
This property is quite useful since it allows us to obtain the $\fmn$ equations of motion directly from the $\gmn$ equations.

Finally, the matter Lagrangian $\Lag_{\rm m}$ contains the SM matter fields $\Phi$, which we have taken to be minimally coupled only to $\gmn$ here. This choice is without loss of generality since the theory treats the metrics symmetrically and, in the bimetric theory, standard matter fields can only couple minimally to one of the tensor fields without introducing ghost instabilities~\cite{Yamashita:2014fga, deRham:2014naa}.

The equations of motion that follow from the action~\eqref{bgaction} are given by,
\begin{subequations}\label{eombim}
\begin{align}
g{\rm-eom}:&\qquad\quad\,\,
\mathcal{G}_{\mu\nu}(g)+\frac{\alpha^2\mpl^2}{1+\alpha^2}\,V_{\mu\nu}(g,f)
=\frac{1+\alpha^2}{\mpl^2}\,T_{\mu\nu}\,, \label{bgeoms_g}\\ \vspace{3cm}
f{\rm-eom}:&\qquad\quad\,\,
\mathcal{G}_{\mu\nu}(f)+\frac{\mpl^2}{1+\alpha^2}\,\tilde{V}_{\mu\nu}(g,f)=0\,.
\label{bgeoms_f}
\end{align}
\end{subequations}
Here $\mathcal{G}_{\mu\nu}(g)=R_{\mu\nu}(g)-\tfrac1{2}\gmn R(g)$ is the Einstein tensor computed with respect to $\gmn$ and $\mathcal{G}_{\mu\nu}(f)=R_{\mu\nu}(f)-\tfrac1{2}\fmn R(f)$ is the Einstein tensor computed with respect to $\fmn$. The interaction contributions $V_{\mu\nu}, \tilde V_{\mu\nu}$ and the matter stress-energy $T_{\mu\nu}$ are defined by,
\be\label{Vmndefs}
V_{\mu\nu}\equiv \frac{-2}{\sqrt{|g|}}\frac{\p(\sqrt{|g|}V)}{\p g^{\mu\nu}}\,,\qquad
\tilde{V}_{\mu\nu}\equiv \frac{-2}{\sqrt{|f|}}\frac{\p(\sqrt{|g|}V)}{\p f^{\mu\nu}}\,,\qquad
T_{\mu\nu}\equiv\frac{-1}{\sqrt{|g|}}\frac{\p(\sqrt{|g|}\Lag_{\rm m})}{\p g^{\mu\nu}}\,.
\ee
As noted, \eqref{bgeoms_f} can be obtained directly from~\eqref{bgeoms_g} by making use of the interchange symmetry \eqref{intsym}. The interaction contributions are matrix polynomials in $S$, which can be written,
\begin{align}\label{Vmndef}
V_{\mu\nu}=g_{\mu\rho}\sum_{n=0}^3(-1)^n\beta_nY^\rho_{(n)\,\nu}(S)\,,\qquad
\tilde{V}_{\mu\nu}=f_{\mu\rho}\sum_{n=0}^3(-1)^n\beta_{4-n}Y^\rho_{(n)\,\nu}(S^{-1})\,,
\end{align}
where the tensors $Y_{(n)}(S)$ are defined as,
\be
Y^\rho_{(n)\,\nu}(S)=\sum_{k=0}^n(-1)^ke_k(S)[S^{n-k}]^\rho_{~\nu}\,.
\ee
For example, written out explicitly we have that,
\begin{align}
V_{\mu\nu}=g_{\mu\rho}\biggl[&\beta_0\delta^\rho_{\nu}-
\beta_1\left(S^\rho_{~\nu}-e_1\delta^\rho_\nu\right)
+\beta_2\left([S^2]^\rho_{~\nu}-e_1S^\rho_{~\nu}+e_2\delta^\rho_\nu\right)\nn\\
&-\beta_3\left([S^3]^\rho_{~\nu}-e_1[S^2]^\rho_{~\nu}+e_2S^\rho_{~\nu}-e_3\delta^\rho_\nu\right)\biggr]\,.
\end{align}
We note that $V_{\mu\nu}$ and $\tilde{V}_{\mu\nu}$ as written in \eqref{Vmndef} are symmetric in their indices, although not manifestly so. This follows from the fact that both $S$ and $S^{-1}$ are symmetric whenever their indices are raised or lowered using either of $\gmn$ or $\fmn$.\footnote{This can be proven either by a formal expansion of the square-root~\cite{Hassan:2012wr} or by matrix manipulations~\cite{Baccetti:2012re}.}

The theory defined by the action~\eqref{bgaction} is generally covariant under the diagonal group of common diffeomorphisms. The fact that the interaction potential is covariant on its own implies the following divergence identities (see e.g.~\cite{Damour:2002ws}),
\be\label{covID_1}
\sqrt{|g|}\,g^{\mu\rho}\nabla_\rho V_{\mu\nu} = - \sqrt{|f|}\,f^{\mu\rho}\tilde{\nabla}_\rho \tilde{V}_{\mu\nu}\,,
\ee
as well as the algebraic identities \cite{Hassan:2014vja} (see also~\cite{Baccetti:2012bk, Volkov:2012zb}),
\be\label{covID_2}
\sqrt{|g|}\,g^{\rho\mu}V_{\mu\nu}+\sqrt{|f|}\,f^{\rho\mu}\tilde{V}_{\mu\nu}-\sqrt{|g|}\,V\delta^\rho_\nu
=0\,,
\ee
where $V$ is the interaction potential \eqref{Vdef} appearing in the action \eqref{bgaction}. 
For a covariantly conserved source, the standard Bianchi identities, $\nabla^\mu\mathcal{G}_{\mu\nu}=0$ and $\tilde\nabla^\mu\mathcal{G}_{\mu\nu}=0$, imply the constraint equations $\nabla^\mu V_{\mu\nu}=0$ and $\tilde\nabla^\mu\tilde V_{\mu\nu}=0$. Due to the identity~\eqref{covID_1}, these are not independent and so in all only give 4 constraints. Apart from that, an additional scalar constraint can be constructed~\cite{Bernard:2015uic} (which was first found in the Hamiltonian formulation~\cite{Hassan:2011zd}). These $4+1$ constraints serve to remove 5 dynamical modes from the $10+10=20$ components of the two tensor fields. The diffeomorphism invariance removes $2\times4=8$ more. Bimetric theory therefore propagates $20-8-4-1=7$ degrees of freedom. As we will see next, when such a split makes physical sense, these degrees of freedom correspond to a massless spin-2 field (2) and a massive spin-2 field (5).

\subsection{Proportional solutions \& mass spectrum}\label{sec:propsols}
An important class of solutions in bimetric theory without any matter sources are the proportional solutions, defined by $\bfmn=c^2\bgmn$. For such an ansatz the Bianchi constraints immediately imply that $c^2$ is a constant. In order to simplify notation we will set $c^2=1$ in what follows. This can be done without any loss of generality by scaling $\fmn$ and properly redefining $\alpha$ along with the $\beta_n$. Such a scaling is possible since we do not couple $\fmn$ to matter in our considerations, which gives rise to a redundancy in the parameter space.

For the proportional ansatz the bimetric vacuum equations reduce to~\cite{Hassan:2012wr},
\begin{subequations}\label{eombimprop}
\begin{align}
g{\rm-eom}:&\qquad
\mathcal{G}_{\mu\nu}(\bar g)+\Lambda_g\,\bgmn=0\,, \label{bgeoms_gprop}\\
f{\rm-eom}:&\qquad
{\mathcal{G}}_{\mu\nu}(\bar g)+\Lambda_f\,\bgmn=0\,,
\label{bgeoms_fprop}
\end{align}
\end{subequations}
with constants,
\begin{subequations}
\begin{align}\label{lambdag}
\Lambda_g&=\frac{\alpha^2\mpl^2}{1+\alpha^2}(\beta_0+3\beta_1+3\beta_2+\beta_3)\,,\\
\Lambda_f&=\frac{\mpl^2}{1+\alpha^2}(\beta_4+3\beta_3+3\beta_2+\beta_1)\,.
\label{lambdaf}
\end{align}
\end{subequations}
Consistency between the equations now requires $\Lambda_g=\Lambda_f\equiv\Lambda$.
Since we have set $c=1$ this relation generically fixes one of the $\beta_n$ parameters.\footnote{Due to the aforementioned freedom of rescaling $\fmn$, this constitutes no loss of generality but fixes a redundant parameter. For general $c$ it would instead provide a fourth order polynomial equation for $c$ which generically determines $c=c(\alpha,\beta_n)$ and thereby fully specifies the solution.} This class of solutions thus corresponds to the maximally symmetric solutions of GR. Flat space solutions with $\Lambda=0$ require fixing one of the $\beta_n$ and whenever we discuss flat backgrounds this will always be implicitly assumed. 

For spacetimes admitting Poincar\'e or (Anti) de Sitter isometries the representation theory of spin-2 fields is well known. It is therefore natural to study perturbations of the proportional solutions. Perturbation theory in bimetric theory is notoriously challenging due to the presence of the square root matrix in the interaction potential and the general problem was only recently resolved~\cite{Bernard:2014bfa,Bernard:2015mkk,Bernard:2015uic}. For the proportional solutions, however, the situation simplifies greatly.

We define linear fluctuations $h, \ell$ around the proportional backgrounds by,
\be
\gmn=\bgmn+\hmn\,,\qquad \fmn=\bgmn+\lmn\,.
\ee
The canonically normalised mass eigenstates are then defined through~\cite{Hassan:2012wr},
\begin{subequations}\label{masseigdef}
\begin{align}
\dG_{\mu\nu}&=\frac{\mpl}{1+\alpha^2}\left(\hmn+\alpha^2\lmn\right)\,,\label{masseigG}\\ 
\dM_{\mu\nu}&=\frac{\alpha\,\mpl}{1+\alpha^2}\left(\lmn-\hmn\right)\,,\label{masseigM}
\end{align}
\end{subequations}
where, for future reference, we also note the inverse relations,
\begin{subequations}\label{inverseh}
\begin{align}
\hmn&=\frac{1}{\mpl}\left(\dG_{\mu\nu}-\alpha\dM_{\mu\nu}\right)\,,\label{invgrel}\\
\lmn&=\frac{1}{\mpl}\left(\dG_{\mu\nu}+\alpha^{-1}\dM_{\mu\nu}\right)\,.
\end{align}
\end{subequations}
The parameter $\alpha$ thus quantifies the mixing between the fluctuations. In terms of the mass eigenstates~\eqref{masseigdef} the quadratic part of the action~\eqref{bgaction} diagonalises into (indices are raised and lowered using $\bgmn$),
\begin{align}\label{bgaction2}
S^{(2)}=\int\td^4x\sqrt{|\bar g|}\,\biggl[&
\Lag^{(2)}_{\rm GR}(\dG)+\Lag^{(2)}_{\rm GR}(\dM)
-\frac{m_{\rm FP}^2}{4}\left(\dM_{\mu\nu}\dM^{\mu\nu}-\dM^2\right)\nn\\
&-\frac{1}{\mpl}\left(\dG_{\mu\nu}-\alpha\dM_{\mu\nu}\right)T^{\mu\nu}\biggr]\,,
\end{align}
where $\Lag^{(2)}_{\rm GR}$ is the quadratic theory obtained from the Einstein-Hilbert action including a cosmological constant, i.e.~$\sqrt{|g|}(R-2\Lambda)$. The detailed expression for this is given in eq.~\eqref{app:Lag2GR}. We have defined the Fierz-Pauli mass of the massive spin-2 field,
\be\label{mfpdef}
\mfp \equiv \sqrt{\beta_1 + 2\beta_2 + \beta_3}\,\mpl \equiv \xi\,\mpl \,.
\ee
Note that our parametrisation implies that the parameters $\beta_1, \beta_2, \beta_3$ are on the order of $\mfp^2/\mpl^2$.

The quadratic theory contains a massless graviton $\dG_{\mu\nu}$, which mediates standard gravitational interactions with Planck mass $\mpl$ and an additional massive spin-2 field $\dM_{\mu\nu}$ with mass $\mfp$. Note that the massive spin-2 field couples to the matter stress-energy and therefore also mediates gravitational interactions but with a coupling $\alpha/\mpl$. 

For small $\alpha$, the matter coupling of the massive field will be suppressed with respect to that of the massless field. One may therefore expect to recover a situation close to linearised GR for small enough $\alpha$. 
On the other hand, as we will see later when studying higher-order interactions, while the massive mode decouples from the SM matter, it does not decouple from gravity in the $\alpha\rightarrow0$ limit. In fact, it continues to gravitate with the exact same strength as SM matter. This makes the massive spin-2 field of bimetric theory an interesting candidate for a DM particle.
Similarly, another way to recover linearised GR (at least at low energies) is to consider large values for $\mfp$, which also decouples the heavy spin-2 field from the matter sector. The following section is dedicated to a detailed discussion of the behaviour of the theory in these two parameter regimes.

\section{Recovering General Relativity}\label{sec:BMtoGR}

As any other modification of GR, bimetric theory containing a second tensor field generically changes the laws of gravity. Since GR is well-tested over a large range of energy regimes, we need to carefully evaluate the observational constraints on bimetric theory and make sure that its predictions do not differ too much from GR. 
In this section we will see that there are two different (but overlapping) regions in the parameter space of bimetric theory for which certain classical solutions for the physical metric approach those of GR. In particular, the cosmological as well as the static spherically symmetric solutions to the bimetric equations of motion both resemble GR in the overlap of these two regions.

\subsection{The GR regimes for physical solutions}\label{sec:grlimit}

The two separate parameter regions which recover GR 
for the physical metric $\gmn$ can be motivated based on the linear theory around proportional backgrounds:
\begin{itemize}
\item[(i)] The more general option is to consider a large hierarchy between the ``Planck masses" of the two metrics, i.e.~$\alpha\ll1$. Physically this corresponds to 
a very feeble coupling of the massive spin-2 field to matter sources, irrespective of its mass. It also implies enhanced self-interactions of the massive field and a large value for the physical Planck mass.
All known solutions of bimetric theory coincide with GR solutions for $\gmn$ in the limit $\alpha\rightarrow 0$.

\item[(ii)] The second option is to take the Fierz-Pauli mass $\mfp$ to be large, typically $\mfp^2\gg\Lambda$, which effectively means that $\xi$ in~\eqref{mfpdef} should obey $\Lambda/\mpl^2\ll\xi^2$. Additionally, since $\xi$ sets the scale of $\mfp$ in units of $\mpl$, we should also require $\xi\ll1$. Regarding the massive spin-2 field as DM will turn out to give much more stringent bounds. In physical terms, we would like to make the massive spin-2 field heavy enough such that it effectively decouples from the low-energy theory. This option presumably recovers GR for $\gmn$ only in the linear regime around the proportional backgrounds, since the notion of $\mfp$ has no clear meaning away from these solutions. Nevertheless, this criterion turns out to be useful also in the context of cosmological solutions.
\end{itemize}
Our analysis in section~\ref{sec:pheno} will reveal that observations favour a combination of both these options. We will therefore discuss the two parameter regions in more detail for two physically important classes of solutions, the static point-source and the cosmological solutions.

\subsection{Static spherically symmetric solutions}\label{sec:pointsol}

Local gravity tests tell us that any theory for gravity inside the Solar System, up to 10$\,\mu$m, must follow the predictions of GR to high precision~\cite{Will:2014kxa} and bimetric theory studied in the context of DM has to pass these tests. 
To approximate modified gravity effects inside the Solar System, one considers static spherically symmetric solutions around a massive source (which would correspond to the Sun). 
The gravitational field computed in this approximation must effectively resemble GR up to the precision available so far. 
Another important aspect of studying spherically symmetric solutions is to fix the value of Newton's constant. 
In modified gravity theories, the value derived in this way can, in principle, differ from the corresponding value obtained in cosmology. 
Upon comparing local and cosmological observations, this may lead to extra constraints on the theory.

Historically, the first attempt to build a massive gravity theory -- Fierz-Pauli massive gravity --
has been rejected precisely because it fails even basic Solar System tests. 
The problem arises because the spin-0 mode of the massive graviton adds an extra (fifth) force to gravitational interactions and does not decouple in the limit of small graviton mass, $m_{\rm FP}\to 0$. This effect is known as vDVZ discontinuity~\cite{vanDam:1970vg,Zakharov:1970cc} and we review it briefly in appendix~\ref{app:mg}.
It was conjectured in~\cite{Vainshtein:1972sx} (and confirmed explicitly much later \cite{Babichev:2009jt,Babichev:2010jd,Babichev:2009us}, see also \cite{Alberte:2010it}) that the inclusion of nonlinear interactions for the spin-2 field cures this problem
and that GR is restored in the limit of small graviton mass. Today this feature is known as the Vainshtein mechanism and it operates in a plethora of modified gravity models, see e.g.~the review~\cite{Babichev:2013usa}.

As we will see below, bimetric theory with a very heavy spin-2 mass does not require the Vainshtein mechanism, 
since the solution is linear all the way down to very small lengths, where gravity is not yet tested. 
In this case, the spherically symmetric solutions recover GR despite being linear, which is in sharp contrast to massive gravity, for which the linear regime always leads to contradictions with Solar System tests. 
This is a consequence of the fact that massive gravity contains only one propagating massive graviton, while bimetric theory has an additional massless graviton. On the other hand, for bimetric parameters which require us to go beyond the linearised approximation, we still have to rely on the Vainshtein mechanism to restore GR. 
In this case, the restoration through the Vainshtein mechanism is quite similar to massive gravity.

\subsubsection{Derivation}\label{sec:sperical}
An appropriate ansatz for spherically symmetric solutions in bimetric theory reads~\cite{Babichev:2013pfa},
\begin{align}\label{ds2g}
ds^2_{(g)} & = -e^\nu dt^2 + e^\lambda dr^2 + r^2 d\Omega^2\,, \\
ds^2_{(f)} & = -e^{\tilde\nu} dt^2 + e^{\tilde\lambda} (r+r\mu)'^2 dr^2 + (r+r\mu)^2 d\Omega^2\,,
\label{ds2f}
\end{align}
where $\nu$, $\lambda$, ${\tilde\nu}$, $\tilde\lambda$ and $\mu$ are functions of $r$.
We will always assume that the functions $(\nu, \lambda, \tilde{\nu}, \tilde\lambda) $ are much smaller than unity,
corresponding to weak sources. On the other hand, the function $\mu$ can be either small or large and in the latter case carries information about nonlinear effects. In massive gravity the function $\mu$ can be associated with a St\"uckelberg field~\cite{Babichev:2009us}.
The common diffeomorphism invariance has been used in the above ansatz to remove a function in front of $r^2 d\Omega^2$ in the 
$\gmn$ metric.

Linearising the equations of motion (\ref{eombim}), 
one can obtain the following solutions~\cite{Babichev:2013pfa} (see also \cite{Comelli:2011wq,Enander:2013kza}),
\begin{equation}
\label{bimulin}
\begin{split}
	\mu &= -\, \frac{C_2(1+\alpha^2) \, e^{-m_{\rm FP}r}
   \left(1 + m_{\rm FP}r+m^2_{\rm FP}r^2\right)}{3\,m^4_{\rm FP} \,r^3} \,,  \\
	\lambda &= \frac{C_1}{r} + \frac{2\,C_2 \, \alpha^2\, e^{-m_{\rm FP}r}\left(1 + m_{\rm FP}r\right)}
	{3\,m^2_{\rm FP} \,r } \,,  
	\quad  \nu = - \frac{C_1}{r} - \frac{4\, C_2 \,\alpha^2(1+\alpha^2)\, e^{-m_{\rm FP}r}  }{3\,m^2_{\rm FP} \,r} \\
   \tilde\lambda &= \frac{C_1}{r} - \frac{2\,C_2 \,  e^{-m_{\rm FP}r}\left(1 + m_{\rm FP}r\right)}
	{3\,m^2_{\rm FP} \,r } \,, \quad
	\tilde\nu = - \frac{C_1}{r} + \frac{4\, C_2\, e^{-m_{\rm FP}r}  }{3\,m^2_{\rm FP} \,r} \,, 
\end{split}
\end{equation}
where $C_1$ and $C_2$ are two integration constants, to be fixed by matching the solution to the source. 
Depending on the parameters of the model and the mass of the central source, 
the linearised approximation may not be valid for all distances.
The above expressions were obtained under the assumption that nonlinearities in $\mu$ can be neglected. 
The Vainshtein mechanism starts to operate exactly when nonlinearities in $\mu$ become important.

The equations of motion (\ref{eombim}) can be solved analytically in a different regime, which does not rely on linearity in $\mu$. 
Assuming that we are deep inside the Compton wavelength, $r\ll m^{-1}_{\rm FP}$, one finds a seventh order algebraic equation for $\mu$ \cite{Babichev:2013pfa}. 
The solution in this regime is valid down to small radii, and it can be matched to a solution inside the source. 
To this end, we introduce the Schwarzschild radius,
\begin{equation}\label{rS}
	r_S = \frac{1+\alpha^2}{\mpl^2}\int^{R_\odot}_0 \rho\, r^2 dr,
\end{equation}
where $\rho$ is the density inside the central source and $R_\odot$ is the radius of the body. 
Note that the above expression has an extra factor $(1+\alpha^2)$, resulting in an extra factor $(1+\alpha^2)$ in Newton's constant  with respect to GR. 
Another relevant scale is the Vainshtein radius, 
\begin{equation}\label{rV}
	r_{\rm V} = \left(\frac{r_S}{m^2_{\rm FP}}\right)^{1/3}\,,
\end{equation}
below which the nonlinearities in $\mu$ kick in. 
One then finds that for $r_V\ll r \ll m^{-1}_{\rm FP}$, 
\begin{equation}\label{insideC}
\begin{split}
\mu &= -\frac{r_S}{3 m^2_{\rm FP} r^3},\;\; \lambda = \frac{(3+2\alpha^2)r_S}{3(1+\alpha^2)r},\;\; \nu=  -\frac{(3+4\alpha^2)r_S}{3(1+\alpha^2)r},\\
\tilde\lambda &= \frac{r_S}{3(1+\alpha^2)r},\;\; \tilde\nu=  -\frac{r_S}{3(1+\alpha^2)r}.
\end{split}
\end{equation}
At smaller radii, $r\ll r_V$ the solution changes its form to,
\begin{equation}
\label{insiderV}
\mu = \text{const},\;\; \lambda = \frac{r_S}{r},\;\; \nu=  -\frac{r_S}{r},\;\; 
\tilde\lambda \propto m^2_{\rm FP} r^2,\;\; \tilde\nu \propto m^2_{\rm FP} r^2,
\end{equation}
which restores GR for the physical metric $\gmn$. This is precisely the Vainshtein mechanism operating for radii $r\ll r_V$.
The constant expression for $\mu$ depends on the parameters of the Lagrangian as well as the exact form of $\tilde\lambda$ and $\tilde\nu$.

The matching of the linearised solution (\ref{bimulin}) and the solution inside the Compton wavelength (\ref{insideC})
fixes the constants of integration 
$C_1$ and $C_2$ as follows,
\begin{equation}\label{C12}
	C_1=\frac{r_S}{1+\alpha^2},\quad C_2 = \frac{m^2_{\rm FP}r_S}{1+\alpha^2}.
\end{equation}
Note that in the linear regime, the metric functions of $\gmn$ receive an extra factor $1/(1+\alpha^2)$ 
with respect to their behaviour in the Vainshtein regime. Moreover, the above matching is only valid when the Vainshtein regime is present at all. 
Otherwise, when the linear regime is valid all the way down to the source, the solutions obtained in the linearised approximation 
must be matched to the source. We discuss this case below, in the context of large values for the spin-2 mass.

It is worth pointing out that the scale of nonlinearity $r_V$ is not directly related to the validity of the perturbative expansion for bimetric theory which we will address in section~\ref{sec:valexp}. The scale where classical solutions become nonlinear depends on an extra scale of the problem, namely the mass of the central source. 

\subsubsection{The region $\mfp^2\gg\Lambda$}
Let us now discuss the region where the Fierz-Pauli mass is large, i.e.~$\mfp^2\gg\Lambda$.
In the limit of infinitely large mass, both the Compton wavelength $\mfp^{-1}$ and the Vainshtein radius $r_V$ vanish. 
This means that the linearised approximation is valid for all radii.
From (\ref{bimulin}) one can easily see that for $\mfp\rightarrow\infty$ we find the GR solution,
\begin{equation}\label{largemfp}
\mu = 0, \quad \lambda = \tilde\lambda = \frac{C_1}r,\quad \nu=\tilde\nu = -\frac{C_1}r\,.
\end{equation} 
Note that, in this case, one cannot use the expressions in (\ref{C12}) for the integration constants since they were obtained by assuming that the Vainshtein regime operates for small radii. 
In the limit $\mfp \to \infty$ the solution always remains linear and one needs to redo the matching to the source. 
We will not go into the details of this computation, but only give the result.
Assuming that $\lambda$ and $\tilde\lambda$ are given by (\ref{largemfp}) outside the source, and by the same expressions, 
but with $C_1$ being a function of the radius inside the source, one obtains,
\begin{equation}\label{ssol}
\lambda = -\nu = \frac{1}{r\, \mpl^2}\int^r_0\rho\, r'^2 dr'\,.
\end{equation}
We conclude that the local Planck mass coincides with our original definition of $\mpl$.

This result could have been anticipated from the action written in the terms of mass eigenstates~(\ref{bgaction2}). 
Since the linear approximation is valid in the limit $\mfp \to \infty$ 
(at least outside the sources), the quadratic action (\ref{bgaction2}) is sufficient for 
studying spherically symmetric solutions.  
Note that both the massless $\dG$ and massive $\dM$ spin-2 field contribute to the physical metric $g_{\mu\nu}$ via the relation~(\ref{invgrel}).
The coupling constant between the massless graviton $\dG_{\mu\nu}$ and the source is precisely $\mpl^{-1}$.
At the same time, the massive spin-2 mode $\dM_{\mu\nu}$ is also excited by the source term, but because of the 
vanishing Compton wavelength $\mfp^{-1}$, the solution for $\dM_{\mu\nu}$ outside the source vanishes in the limit $\mfp \to \infty$. 
As a consequence, the only contribution to the physical metric $h_{\mu\nu}$ comes from the massless mode $\dG_{\mu\nu}$, c.f.~(\ref{invgrel}).
Hence we recover exactly the results in (\ref{largemfp}) and (\ref{ssol}).

\subsubsection{The region $\alpha\ll1$}

Now we turn to the limit $\alpha\to 0$, where at the same time we keep $\mfp$ constant. 
Neither the Vainshtein radius in (\ref{rV}) nor the Compton wavelength $\mfp^{-1}$ vanish in this limit.
Therefore, in contrast to the case $\mfp\to \infty$ considered above, the theory enters a nonlinear regime 
for small enough distances (at least for large enough $r_S$).
According to the general discussion in section~\ref{sec:sperical}, 
the solution in the linear regime is valid for $r\gg r_V$. From (\ref{bimulin})
 in the $\alpha\to 0$ limit we then find, 
\begin{equation}\label{smalla}
	\lambda = -\nu = \frac{r_S}r\,,
\end{equation}
where we also used (\ref{C12}). This shows that, for large radii, GR is restored. 
The physical Planck mass is again $\mpl$ since from (\ref{rS}) with $\alpha\to 0$ we get,
\begin{equation}\label{rSsmalla}
	r_S = \frac{1}{\mpl^2}\int^{R_\odot}_0 \rho\, r^2 dr\,.
\end{equation}
Notice that, in contrast to the case of large mass, the function $\mu$ (the ``St\"uckelberg field") does not vanish, 
\begin{equation}
	\mu = \frac{r_S \, e^{-m_{\rm FP}r}  \left(1 + m_{\rm FP}r+m^2_{\rm FP}r^2\right)}{3\,m^2_{\rm FP} \,r^3} \,.
\end{equation}
At $r\sim r_V$ it becomes of order unity, confirming that the linear approximation breaks down at this scale. 
Thus we have to resort to the nonlinear regime for $r < r_V$.
Using (\ref{insideC}) and (\ref{insiderV}), it is straightforward to show that, in the limit $\alpha\to 0$, the solution in the nonlinear regime with $r< r_V$ is again given by (\ref{smalla}) and thus coincides with GR. 

Once more we could have started from the quadratic action (\ref{bgaction2}) and anticipated part of the result.
Solving the equations for the massless eigenstate $\dG_{\mu\nu}$, we recover the GR solution as long as the linear regime is valid. 
Newton's constant is given in terms of the Planck mass $\mpl$, as can be read off from (\ref{bgaction2}). 
The relation (\ref{invgrel}) shows that the only contribution to $h_{\mu\nu}$ is $\dG_{\mu\nu}$ in the limit $\alpha\to 0$ 
and hence $h_{\mu\nu}$ in the linear regime corresponds to a GR solution.
This is in accordance with (\ref{smalla}), which was obtained from the general formalism of spherically symmetric solutions. 
On the other hand, when nonlinearities kick in, we cannot rely anymore on the quadratic action (\ref{bgaction2}) since the nonlinear terms become as important as the linear ones. In other words, (\ref{bgaction2}) is not sufficient for studying the behaviour of the metric inside the Vainshtein radius. In this regime GR is restored by nonlinear effects, c.f.~(\ref{insiderV}), which is independent of taking any parameter limit.

\begin{figure}[thbp]
\begin{center}
\includegraphics[width=0.8\textwidth]{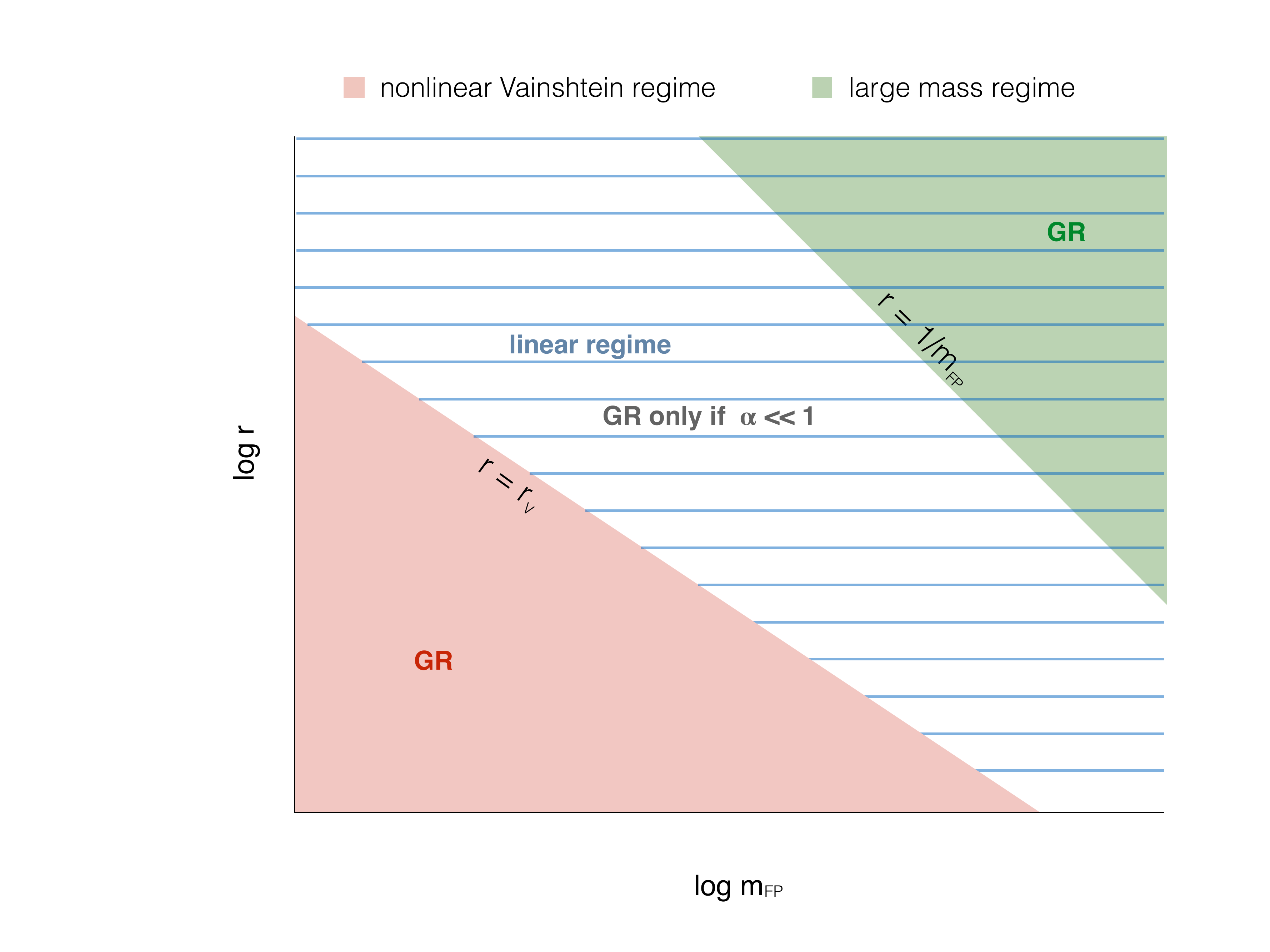}
\end{center}
\caption{Schematic diagram showing how GR is restored for different distance scales $r$, depending on the spin-2 mass $\mfp$. In the red-shaded region, the solution is nonlinear and GR is recovered via the Vainshtein mechanism. In the linear regime (blue-striped region), for $r<\mfp^{-1}$, it is necessary to require $\alpha\ll 1$ in order to recover GR (white region), while for $r\gg\mfp^{-1}$, GR is recovered due to the exponential fall-off of the Yukawa potential (green-shaded region).}
\label{regimes}
\end{figure}

We conclude that in the limit of small $\alpha$, GR is restored for all radii, with Newton's constant given by the Planck mass $\mpl$.
It is worth emphasising that this type of GR restoration for the physical metric $\gmn$ is quite nontrivial, since it involves a transition between the linear and nonlinear regimes. 
All features defining the regime of the solution are hidden in the second metric $\fmn$, including the St\"uckelberg field $\mu$. 
The different regimes in which GR is recovered for static spherically symmetric solutions are visualised in Fig.~\ref{regimes}.

\subsection{Cosmological solutions}\label{sec:cosmosol}

Just as the local gravity tests, the GR based $\Lambda$CDM concordance model has been confirmed to high precision and therefore puts stringent constraint on modifications of gravity. 
The homogeneous and isotropic solutions to the bimetric equations of motion were first derived in Ref.~\cite{Volkov:2011an, vonStrauss:2011mq, Comelli:2011zm}. For general parameters they give rise cosmological observables which differ significantly from GR predictions. 
The behaviour of cosmological solutions for large $\mfp$ has previously been discussed in~\cite{DeFelice:2014nja} and for small $\alpha$ in~\cite{Akrami:2015qga}. Here we will take a slightly different approach with respect to these references, which will allow us to treat both cases simultaneously and to show that bimetric theory again resembles GR in the overlap of these parameter regions. 
For a review of bimetric cosmology, we refer the reader to~\cite{Solomon:2015hja}.

\subsubsection{Derivation}
Restricting our analysis to the bidiagonal case, we can put the metrics on the form, 
\begin{align}
\gmn\td x^\mu\td x^\nu&=-\td t^2+a^2(t)\left(\frac{\td r^2}{1-kr}+r^2\td\Omega^2\right)\,,\\
\fmn\td x^\mu\td x^\nu&=-X^2(t)\td t^2+Y^2(t)\left(\frac{\td r^2}{1-kr}+r^2\td\Omega^2\right)\,,
\end{align}
where $k=0, \pm 1$ corresponds to a flat, open and closed universe respectively. It follows from the equations of motion that we must have the same $k$ in both metrics. Here all isometries have been used to put $\gmn$ on the standard FLRW form. The ``lapse" function $X(t)$ can be solved for directly from the Bianchi constraint to give $X=\dot Y/\dot a$.

The matter source coupled to $\gmn$ is taken to be a perfect fluid, $T^{\mu}_{~\nu}=\mathrm{diag}(-\rho, p, p, p)$. The spatial scale factors $a(t)$ of $\gmn$ and $Y(t)$ of $\fmn$ are then solutions to the dynamical equation, with $r(t)\equiv Y/a$,
\be\label{friedmannmod}
H^2+\frac{k}{a^2}
=(1+\alpha^2)\frac{\rho}{3\mpl^2}
+\frac{\alpha^2\mpl^2}{1+\alpha^2}\left(\frac{\beta_0}{3}+\beta_1\,r+\beta_2\,r^2
+\frac{\beta_3}{3}\,r^3\right)\,,
\ee
and the algebraic constraint,
\be\label{cosmoconstr}
\frac{\alpha^2\beta_3}{3} \,r^4
+\left(\alpha^2\beta_2-\frac{\beta_4}{3}\right)\,r^3
+\left(\alpha^2\beta_1-\beta_3\right)\,r^2
+\left(\frac{(1+\alpha^2)^2\rho}{3\mpl^4}+\frac{\alpha^2\beta_0}{3}-\beta_2\right)\,r
-\frac{\beta_1}{3}=0\,.
\ee
The first of these is a modified Friedmann equation for the physical scale factor $a(t)$ while the second determines $r(t)$ in terms of $\rho(t)$. In addition to the linear dependence on the energy density $\rho$, the squared Hubble function $H^2=(\dot{a}/a)^2$ is now also sourced by the contribution coming from the interaction potential. In general, one solves the polynomial equation \eqref{cosmoconstr} for $r$ in terms of $\rho$ and plugs the solution back into \eqref{friedmannmod}. This results in an equation of the form $H^2=F[\rho]$, where $F$ is an analytic function of the energy density whose precise form is determined by the choice of bimetric parameters. Generically, the nonlinear nature of $F[\rho]$ leads to significant deviations of bimetric cosmology from $\Lambda$CDM.

In addition to the above equations, the matter source is subject to the same continuity equation as in GR. The various source components in $\rho(t)$ therefore dilute in time in the standard way. As matter dilutes and $\rho(t)\rightarrow 0$, we see from the algebraic equation~\eqref{cosmoconstr} that $r(t)\rightarrow\,\,$const.~and thus the metrics become proportional with $r^2$ being the constant of proportionality. Moreover, it follows from~\eqref{friedmannmod} that this late-time de Sitter attractor solution has a cosmological constant given by,
\be\label{Lambdarel}
\Lambda=\frac{\alpha^2\mpl^2}{1+\alpha^2}(\beta_0+3\beta_1+3\beta_2+\beta_3)
=\frac{\mpl^2}{1+\alpha^2}(\beta_4+3\beta_3+3\beta_2+\beta_1)\,.
\ee
Here we have taken the asymptotic constant value of $r$ to be $r=1$ without any loss of generality (c.f.~our discussion in beginning of section~\ref{sec:propsols}). The fact that there is a de Sitter attractor at late times is crucial for our logic of extracting phenomenology out of these solutions since observations suggest that we are now living in a cosmological epoch dominated by a cosmological constant.

From the algebraic equation~\eqref{cosmoconstr} we can immediately infer that as soon as the matter density obeys $\rho(t)\ll\beta_i\,\mpl^4$ we can neglect its contribution and solve for $r=\,\,$const.~up to small corrections. To quantify these corrections we make a power series ansatz,\footnote{It follows from the analytic implicit function theorem that $r(\rho)$ is indeed analytic around the solution with $r=\,\,$const., provided that $2\Lambda-3\mfp^2\neq0$.\label{foot:PM}}
\be
r(\rho)=1+\sum_{n\geq1}a_n\left(\frac{\rho}{\mpl^4}\right)^n\,,
\ee
which we plug into~\eqref{cosmoconstr} and subsequently solve for the coefficients $a_n$. After plugging the resulting expression for $r(\rho)$ back into~\eqref{friedmannmod}, the modified Friedmann equation reads,\footnote{In~\cite{Aoki:2013joa} a similar expansion was considered in a bimetric setup with (twin) matter fields coupled also to $\fmn$. That work only considered the first correction to the Friedmann equation, i.e.~the correction to the constant multiplying the term linear in $\rho$.}
\begin{align}\label{expandedFriedmann}
H^2+\frac{k}{a^2}=&\,\frac{\Lambda}{3}
+\frac{\rho}{3\mpl^2}\left[1-\frac{2\alpha^2\left(\Lambda/\mfp^2\right)}{3-2\left(\Lambda/\mfp^2\right)}\right]
+\frac{\rho^2}{\mpl^2\mfp^4}\frac{\alpha^2(1+\alpha^2)(\beta_1-\beta_3)\left(\Lambda/\mfp^2\right)}{\left(3-2\left(\Lambda/\mfp^2\right)\right)^3}\nn\\
&+\frac{\rho^3}{\mpl^4\mfp^6}\frac{\alpha^2(1+\alpha^2)}{3\left(3-2\left(\Lambda/\mfp^2\right)\right)^5}\biggl[
9(\beta_1-\beta_3)+3\left((1+3\alpha^2)\beta_1+(1-\alpha^2)\beta_3\right)\frac{\Lambda}{\mfp^2}\nn\\
&\qquad\qquad\qquad
-9(1+\alpha^2)(\beta_1-\beta_3)^2\frac{\mpl^2\Lambda}{\mfp^4}
-2(1+\alpha^2)(3\beta_1-\beta_3)\frac{\Lambda^2}{\mfp^4}
\biggr]+\dots
\end{align}
Before discussing the validity of this expansion and what it implies for the parameters, we make some general remarks. First of all, we note that all but the first correction vanish for $\beta_1=0=\beta_3$. As can be seen directly from~\eqref{cosmoconstr}, in this case it is easy to obtain an exact solution which effectively result in a cosmological constant and a modified Planck mass, matching the first two terms in the right-hand side above. Secondly, we note that the expansion breaks down when $3\mfp^2=2\Lambda$. This value saturates the so-called Higuchi bound, $3\mfp^2\geq2\Lambda$, which is a well-known unitarity bound for massive spin-2 fields propagating in de Sitter spacetime~\cite{Higuchi:1986py, Higuchi:1989gz}. At the point of saturation a linear gauge symmetry makes the helicity-0 mode of the massive spin-2 field non-dynamical~\cite{Deser:1983mm, Deser:2001us}. The quest for a nonlinear realisation of this linear gauge symmetry has received a lot of attention lately and the phenomenon has also been studied within the bimetric framework~\cite{Hassan:2012gz, Hassan:2012rq, Hassan:2013pca, Hassan:2015tba}. Although this is a very interesting point in the bimetric parameter space, here we will mainly focus on the regime $\mfp^2\gg\Lambda$.

As for the validity of the expansion, we note that at $n$th order in $\rho$ the most dominant correction to the GR term $\rho/3\mpl^2$ comes with a prefactor on the order of,
\beqn
\alpha^2\left(\frac{\rho}{\mpl^2\mfp^2}\right)^{n-1}\,.
\eeqn
In order to see this, recall the definition of the Fierz-Pauli mass~(\ref{mfpdef}) which implies that $\beta_n\mpl^2\sim \mfp^2$ for $n=1,2,3$. Less relevant contributions to the $n$th order in $\rho$ are suppressed by additional factors of $\alpha^2$ and/or by powers of $\Lambda/\mfp^2$ 

We conclude that a higher order term in $\rho$ is generically smaller than a lower order term if $\rho\lesssim \mpl^2\mfp^2$. At the present cosmological epoch this is of course quite easy to satisfy even for a tiny mass, since presently $\rho/\mpl^2\approx H_0^2\approx 10^{-84}\,\GeV^2$. Thus, if $\mfp\gg H_0$ (or equivalently $\mfp^2\gg\Lambda$) we may safely use the expansion to estimate deviations from GR. In order to get a rough order of magnitude estimate for when the validity may break down, we use the fact that at early times we may relate the energy density to the temperature via $\rho\approx T^4$. The bound then implies validity of the expansion for temperatures $T\lesssim 10^9\,\GeV\times \left(\mfp/\GeV\right)^{1/2}$. In practice, this bound will be even less stringent due to the additional $\alpha^2$ suppression.

\subsubsection{The region $\mfp^2\gg\Lambda$}\label{sec:cosmolargem}

Interestingly, despite being physically well-defined only for the proportional solutions, the parameter combinations $\Lambda$ and $\mfp$ play a major role in the expansion~\eqref{expandedFriedmann}. We see that, apart from the pure cosmological constant term, $\Lambda$ always enters via the dimensionless ratio $\Lambda/\mfp^2$. In particular, the lowest order correction to GR in~\eqref{expandedFriedmann} comes as a renormalisation of the physical Planck mass $\mpl$ and is proportional to $\alpha^2\left(\Lambda/\mfp^2\right)$. 
The strongest bounds on the value of the Planck mass obtained from cosmological/large scale considerations comes from Big Bang Nucleosynthesis (BBN) observations (see e.g.~\cite{Iocco:2008va}). However, these constrain the value of the physical Planck mass only to within about 10\%, or at best a few percent. Hence, this constraint alone does not require a very large $\mfp$ or a small $\alpha$, since the combination $\alpha^2\left(\Lambda/\mfp^2\right)$ only has to be less than $\sim0.1$.

A stronger motivation for considering $\mfp^2\gg\Lambda$ comes from the work~\cite{DeFelice:2014nja}, which considered perturbations of the cosmological solutions discussed above. It turns out that in general a gradient instability is present in the scalar sector which threatens to invalidate linear perturbation theory~\cite{Comelli:2012db, Konnig:2014dna, Lagos:2014lca}. This instability however disappears when $(1+\alpha^2)\Lambda\ll\mfp^2$ and $(1+\alpha^2)\rho\ll2\alpha^2\mpl^4$, provided a mild bound on the parameters is satisfied (to wit,~$\beta_2+\beta_3>0$). Therefore, considering a large mass is a safe way of ensuring that standard techniques of perturbation theory are still applicable.

From the expansion~\eqref{expandedFriedmann} it is also clear that the condition $\mfp^2\gg\Lambda$ alone does not affect all of the corrections to pure GR like behaviour and therefore it cannot serve to fully recover GR from bimetric theory. However, taken together with $\alpha\ll1$, deviations of the cosmological solutions from GR are small at all orders in the expansion. As we will see later, if we treat the massive spin-2 field as a DM candidate, then phenomenology indeed favours the region where $\mfp^2\gg\Lambda$ and $\alpha\ll1$. This ensures the compatibility of our model with cosmological observations.

In the literature on bimetric cosmology it is customary to consider a very small mass for the massive spin-2 field, namely $\mfp^2\sim\Lambda$. This is contrary to our approach, but let us briefly comment on the two main motivations for considering a small mass. 
One is partly historical, relying on intuition from massive gravity where the Vainshtein mechanism is responsible for recovery of GR like behaviour. As we saw in section~\ref{sec:pointsol}, this reasoning is not valid within bimetric theory since, in fact, GR is recovered for large values of $\mfp^2$ without invoking the Vainshtein mechanism. The second motivation is that a small value of the spin-2 mass could lead to a small self-acceleration scale which is ``technically natural". This is based on the argument that a vanishing mass restores the full diffeomorphisms and therefore a small value of the mass is protected by a symmetry from receiving large quantum corrections. 
One may object to this naturalness argument on the ground that i) it perhaps too na\"ively carries results from global symmetries over to local symmetries, ii) gravity seems so far to be exempt from obeying any naturalness criterion and  iii) there is no fundamental reason to expect naturalness to be a sufficient guide. 
In addition to these objections, if $\mfp^2\sim\Lambda$, the BBN constraints may actually be more worrisome phenomenologically unless one also requires a small value for $\alpha$.

We shall not dwell on this issue further but merely note that, if we take the bimetric theory seriously as a model of gravitational interactions, then it certainly seems favourable both from a theoretical and phenomenological perspective to accept a large mass of the additional spin-2 field. Note that this requires us to fine tune the combination of $\beta_n$ parameters in~(\ref{Lambdarel}), in order to produce a small value for the scale of cosmological acceleration.

\subsubsection{The region $\alpha\ll1$}\label{sec:smallalpha}

We stress from the onset that~\eqref{expandedFriedmann} is not an expansion in $\alpha$. Nevertheless, all corrections to the lowest order terms which resemble GR come with at least an $\alpha^2$ suppression, such that a small $\alpha$ leads to a GR like behaviour of the solution. In other words, although $\alpha$ does not always come into play when comparing higher order terms in the expansion, it does affect the relation between the corrections with respect to the lowest order GR like terms. Note that this is of course already obvious from the nonlinear $\gmn$ equations (\ref{bgeoms_g}) and the modified Friedmann equation (\ref{friedmannmod}).

In fact, the possibility to restore GR in the quadratic action by taking $\alpha\rightarrow 0$ generalises to the full nonlinear level, also beyond the cosmological solutions. To see this, let us recall the bimetric equations of motion given in~\eqref{eombim}. For small $\alpha$ the $\fmn$ equations take the form
\be
\mathcal{G}_{\mu\nu}(f)+\mpl^2\,\tilde{V}_{\mu\nu}(g,f)
+\mathcal{O}(\alpha^2)=0\,,
\label{bgeoms2_f}
\ee
where the $\mathcal{O}(\alpha^2)$ corrections simply comes from expanding the factor $1/(1+\alpha^2)$ in front of $\tilde V_{\mu\nu}$ and these can therefore safely be neglected. Now, for regimes where the curvature satisfies $R(f)\ll\beta_i\mpl^2\sim \mfp^2$, we can neglect also the kinetic term.\footnote{From the relation~\eqref{Lambdarel} we infer that some $\beta_i$ may scale as $\alpha^2$. In that case the condition on the curvature may turn into $R(f)\ll\alpha^2\mpl^2$. Later, in section~\ref{sec:heavymass}, we will indeed restrict ourselves to energies satisfying $E\ll\alpha\,\mpl$.} In this case the $\fmn$ equations imply, to first order, that $\fmn$ solves the algebraic equation $\tilde V_{\mu\nu}$=0. The generic solution to the $\fmn$ equation for small $\alpha$ is then that the metrics are nearly proportional. 
As a consequence, the $\gmn$ equations assume the form
\be
\mathcal{G}_{\mu\nu}(g)+\Lambda\,\gmn=\frac{1}{\mpl^2}T_{\mu\nu}
+\mathcal{O}\big(\alpha^2, R(f)/(\beta_i\mpl^2)\big)\,.
\label{bgeoms2_g}
\ee
This is consistent with what we found from the cosmological solutions and supports the fact that all solutions for $\gmn$ approach GR like solutions for small enough $\alpha$ in the energy regimes where $R(f)\ll\beta_i\mpl^2\sim\mfp^2$.

One may expect that, for small but non-vanishing~$\alpha$, all new effects introduced by the presence of the massive spin-2 mode come in as corrections of $\mathcal{O}(\alpha^2)$. Interestingly, this turns out not to be the case. For instance, even in the exact $\alpha\rightarrow 0$ limit, the bimetric interaction potential contributes to the effective cosmological constant $\Lambda$ in~\eqref{expandedFriedmann} and \eqref{bgeoms2_g}, giving rise to background curvature even in the absence of matter. This shows that the universe in bimetric theory can be self-accelerating, i.e.~have $\ddot{a}>0$, even in the absence of vacuum energy (i.e.~for $\beta_0=0$).

\section{Heavy spin-2 field coupled to gravity}\label{sec:heavymass}

The arguments of the previous section, which were based on the cosmological and static point-source solutions, motivate us to further consider the physical implications a heavy spin-2 field coupled to gravity via the ghost-free bimetric interactions. It turns out that such a field naturally has all the desired properties of a suitable DM candidate. In order to elucidate this, we will expand the bimetric action~\eqref{bgaction} in terms of the mass eigenstates defined in~\eqref{masseigdef}. Before considering the explicit form of this expansion, we discuss some of its general features which are independent of the actual form of the ghost-free interactions. Of course, in the end, we will only consider the specific interactions which are free of the Boulware-Deser ghost.

The discussion of section~\ref{sec:nonlinearexpansion} is quite technical and mostly serves to clarify the structure of the expansion and to support its validity. The reader more interested in the final results and their physical interpretation may skip ahead to section~\ref{sec:bmcubic}.

\subsection{General features}\label{sec:nonlinearexpansion}

In order to facilitate the subsequent discussion, we first note several useful relations. To this end let us recall the bimetric action~\eqref{bgaction}, written here without a matter source and with $m^2=\alpha^2m_g^2$,
\begin{align}\label{bgaction3}
S=m_g^2\int\td^4x\biggl[&\sqrt{|g|}R(g)+\alpha^2\sqrt{|f|}R(f)
-2m^2\sqrt{|g|}\,V\left(g^{-1}f\right)\biggr]\,.
\end{align}
This gives rise to the vacuum equations of motion~\eqref{eombim},
\be\label{eombim3}
\mathcal{G}_{\mu\nu}(g)+m^2\,V_{\mu\nu}(g,f)
=0\,, \qquad
\mathcal{G}_{\mu\nu}(f)+\frac{m^2}{\alpha^2}\,\tilde{V}_{\mu\nu}(g,f)=0\,.
\ee
In what follows we will think of the full nonlinear bimetric action as an infinite expansion around the maximally symmetric background solutions $\fmn=\gmn\equiv\bgmn$. It is clear that, absent matter sources, such solutions always exist and imply the background condition $\Lambda_g=\Lambda_f=\Lambda$ with,
\be\label{VpertLambda}
\Lambda_g\,\bgmn
=\left[\frac{-2m^2}{\sqrt{|g|}}\frac{\p(\sqrt{|g|V})}{\p g^{\mu\nu}}\right]_{f=g=\bar g}\,,\qquad
\Lambda_f\,\bgmn
=\left[\frac{-2m^2}{\alpha^2\sqrt{|f|}}\frac{\p(\sqrt{|g|V})}{\p f^{\mu\nu}}\right]_{f=g=\bar g}\,.
\ee
The background condition can therefore, in general, be written,
\be\label{bgcondV}
\alpha^2\frac{\p(\sqrt{|g|V})}{\p g^{\mu\nu}}\biggr|_{f=g=\bar g}
=\frac{\p(\sqrt{|g|V})}{\p f^{\mu\nu}}\biggr|_{f=g=\bar g}\,.
\ee
The general identity~\eqref{covID_2} can then be seen to imply,
\be\label{Vpropsol}
m^2\left[\sqrt{|g|}V\right]_{f=g=\bar g}=\left(\Lambda_g+\alpha^2\Lambda_f\right)\sqrt{|\bar g|}
=(1+\alpha^2)\,\Lambda\,\sqrt{|\bar g|}\,.
\ee
So far, this discussion has been completely general and the above expressions hold for any covariant interaction potential $V$. Let us now restrict our attention to $V$ such that the mass term in the quadratic theory reduces to the Fierz-Pauli one. One can then show that, for fluctuations defined by $\hmn=\gmn-\bgmn$ and $\lmn=\fmn-\bgmn$, the quadratic theory is always diagonalised in terms of the following canonically normalised mass eigenstates\footnote{This follows from the fact that the interactions must depend on $g^{-1}f$ and, to quadratic order we have that $g^{-1}f=\mathbb{1}+\bar g^{-1}(\ell-h)-\bar g^{-1}h\bar g^{-1}(\ell-h)$. Hence the structure of the massive fluctuation is fixed. This in turn fixes also the massless fluctuation.}
\be
\dG_{\mu\nu}=\frac{m_g}{\sqrt{1+\alpha^2}}\left(\hmn+\alpha^2\lmn\right)\,,\qquad
\dM_{\mu\nu}=\frac{\alpha\, m_g}{\sqrt{1+\alpha^2}}\left(\lmn-\hmn\right)\,.
\ee
These are of course consistent with our definitions in~\eqref{masseigG} and~\eqref{masseigM}, which can be seen by using the definition $\mpl=m_g\sqrt{1+\alpha^2}$.

\subsubsection{Mass eigenstates beyond the quadratic expansion}\label{sec:masseigen}

The inverse relations between the metric fluctuations and linear mass eigenstates read,
\beqn\label{invrelme}
\hmn=\frac{1}{\mpl}\left(\delta G_{\mu\nu}-\alpha\delta M_{\mu\nu}\right)\,,\qquad
\lmn=\frac{1}{\mpl}\left(\delta G_{\mu\nu}+\alpha^{-1}\delta M_{\mu\nu}\right)\,.
\eeqn
These are valid as linear field redefinitions which diagonalise the quadratic action. When going to higher orders in perturbation theory, one could in principle consider nonlinear corrections to these relations, e.g.~add terms of order $\delta G^2$ or $\delta M^2$ to the right-hand side of~\eqref{invrelme}. In other words, beyond quadratic level around the maximally symmetric backgrounds, the definition of mass eigenstates becomes ambiguous.\footnote{Going further into the details of this ambiguity is beyond the scope of this work. Possible definitions for nonlinear extensions of the mass eigenstates have been proposed and discussed in~\cite{Hassan:2012wr}.}
However, in the following we will see that the structure of nonlinear interactions justifies the use of the linear relations~\eqref{invrelme} even at higher orders in perturbation theory.

The form of the relations~\eqref{invrelme} implies that we can write the full metrics $\gmn=\bar{g}_{\mu\nu}+\hmn$ and $\fmn=\bar{g}_{\mu\nu}+\lmn$ as follows,
\beqn\label{invrelme2}
\gmn = G_{\mu\nu}-\frac{\alpha}{\mpl}\delta M_{\mu\nu}\,,\qquad
\fmn=G_{\mu\nu}+\frac{1}{\alpha\,\mpl}\delta M_{\mu\nu}\,,
\eeqn
where we have defined a new ``background" metric,
\beqn
G_{\mu\nu}=\bar{g}_{\mu\nu}+\frac{1}{\mpl}\delta G_{\mu\nu}\,.
\eeqn
This structure already hints towards the fact that it may make sense to consider the metric $G_{\mu\nu}$ as a massless field which defines the geometry in which $\dM_{\mu\nu}$ propagates. 

We now consider the full nonlinear bimetric action as an infinite expansion around the solution $\gmn=\fmn=\bar{g}_{\mu\nu}$. We replace the fluctuations $\hmn$ and $\lmn$ around these backgrounds by the linear mass eigenstates using~\eqref{invrelme2}. 
Let us start by focussing on the terms involving only $G_{\mu\nu}$, but no $\delta M_{\mu\nu}$.
Neglecting the massive mode, we effectively deal with the nonlinear bimetric action expanded in
\beqn\label{dmnegl}
\left.\gmn\right|_{\delta M=0}=G_{\mu\nu}=\bar{g}_{\mu\nu}+\frac{1}{\mpl}\delta G_{\mu\nu}\,,\qquad
\left.\fmn\right|_{\delta M=0}=G_{\mu\nu}=\bar{g}_{\mu\nu}+\frac{1}{\mpl}\delta G_{\mu\nu}\,.
\eeqn
Clearly, the Einstein-Hilbert terms as well as the $\beta_0$ and $\beta_4$ terms will just give rise to terms that exactly resemble GR in terms of $G_{\mu\nu}$. The relative factor of $\alpha^2$ between the kinetic terms serve to give an overall factor $\mpl^2$ for this part of the action. The only differences, as compared to GR, could come from the potential which is a function of $g^{-1}f$. But due to \eqref{dmnegl} we have that the terms without $\delta M_{\mu\nu}$ in this matrix reduce to,
\beqn
\left.g^{-1}f\right|_{\delta M=0}=G^{-1}G=\mathbb{1}\,,
\eeqn
and hence (cf.~\eqref{Vpropsol}),
\beqn
\left.m_g^2m^2\sqrt{|g|}\,V(g^{-1}f)\right|_{\delta M=0}=m_g^2(\Lambda_g+\alpha^2\Lambda_f)\sqrt{|G|}=\mpl^2\Lambda\sqrt{|G|}\,,
\eeqn
which again is just a cosmological constant term for $G_{\mu\nu}=\bar{g}_{\mu\nu}+\frac{1}{\mpl}\delta G_{\mu\nu}$. Note that here we have used the background equation $\Lambda_g=\Lambda_f=\Lambda$. Thus, the pure nonlinear self-interactions for the linear massless mode $\delta G_{\mu\nu}$ are nothing but the Einstein-Hilbert term,
\beqn
\left.S(g,f)\right|_{\delta M=0}=\mpl^2\int\dd^4x\,\sqrt{|G|}\,\big(R(G)-2\Lambda\big)\,.
\eeqn
In this sense, $\delta G_{\mu\nu}$ behaves exactly like a massless spin-2 field even in its nonlinear self-interactions.\\

Next, we turn to the non-minimal couplings to the massive field, where it is particularly interesting to study the terms linear in $\delta M_{\mu\nu}$. We formally expand the action to linear order in $\delta M_{\mu\nu}$, treating $G_{\mu\nu}$ as a background metric. This is not a valid perturbative expansion in general because we are not keeping all terms of the same order in $\mpl^{-1}$. Nevertheless, we can look at the terms linear in $\delta M_{\mu\nu}$ and make formal statements about them which will be valid to all orders in the expansion.

In fact, terms involving $\delta M_{\mu\nu}$ only linearly cancel between the two Einstein-Hilbert terms, $\sqrt{g}\,R(g)$ and $\alpha^2\sqrt{f}\,R(f)$. This directly follows from the fact that the massive fluctuation appears in $\hmn$ and $\lmn$ with a relative factor of $\alpha^2$ and opposite sign.

For the potential, we simply Taylor expand to linear order in $\dM$ in the following way,
\begin{align}
m^2\sqrt{|g|}V&=m^2\left[\sqrt{|g|}V\right]_{f=g=G}
+m^2\left[\frac{\p(\sqrt{|g|}V)}{\p g^{\mu\nu}}\frac{\p g^{\mu\nu}}{\p \dM_{\rho\sigma}}
+\frac{\p(\sqrt{|g|}V)}{\p f^{\mu\nu}}\frac{\p f^{\mu\nu}}{\p \dM_{\rho\sigma}}\right]_{f=g=G}\delta M_{\rho\sigma}\nn\\
&=(1+\alpha^2)\Lambda\sqrt{|G|}+\frac{\alpha}{2\mpl}\left(\Lambda_f-\Lambda_g\right)
\sqrt{|G|}\,G^{\mu\nu}\dM_{\mu\nu}\nn\\
&=(1+\alpha^2)\Lambda\sqrt{|G|}\,.
\end{align}
To get to the second line we have used~\eqref{Vpropsol} and~\eqref{VpertLambda} together with the definitions of the fluctuations in~\eqref{invrelme}. The third line then follows from the background relation $\Lambda_g=\Lambda_f$. This demonstrates that the bimetric action in vacuum does not contain any terms linear in the massive fluctuations and, in particular, there is no decay into massless gravitons. These general arguments have also been confirmed from explicit calculations of the cubic and quartic interaction vertices using the ghost-free interactions, as we will see in section~\ref{sec:bmcubic}.
Terms of higher order in $\delta M_{\mu\nu}$ generically do not vanish and give rise to self-interactions of the massive field as well as its non-minimal couplings to the massless graviton $\delta G_{\mu\nu}$. 

The two main conclusions of this section are as follows.
\begin{itemize}
\item[(i)] The nonlinear self-interactions of the massless eigenstate sum up to the standard Einstein-Hilbert action with a cosmological constant. This is consistent with the interpretation of $\dG_{\mu\nu}$ as a massless spin-2 field.

\item[(ii)] There are no terms linear in $\dM_{\mu\nu}$ present in the expansion. This implies that there is no decay of $\dM_{\mu\nu}$ into massless gravitons $\dG_{\mu\nu}$ at tree level.
\end{itemize}
These conclusions are very general in the following sense: They are independent of the form of the interactions apart from reproducing Fierz-Pauli theory at the quadratic level and being covariant at the nonlinear level. They are also independent of any nonlinear field redefinitions of the fluctuations. The physical interpretation of the linear mass eigenstates thus seems to make sense even at the nonlinear level and there does not seem to exist any motivation for considering higher-order corrections to~\eqref{invrelme}.

\subsubsection{Validity of perturbative expansion \& absence of strong coupling}\label{sec:valexp}

We now elaborate in some detail on the validity of the infinite perturbative expansion of the action with $\alpha\ll 1$ in terms of the mass eigenstates and the related issue of strong coupling. 

A general vertex of the schematic form $h^k\,\ell^{\,n}$ in the perturbative expansion of the bimetric interactions around equal backgrounds gives, schematically,
\beqn\label{GMvertices}
h^k\,\ell^{\,n}\sim\sum_{s=0}^k\sum_{r=0}^n
\frac{\alpha^{s-r}}{\mpl^{k+n}} \delta G^{k+n-s-r}\delta M^{s+r}\,,
\eeqn
where we have suppressed the index structure along with numerical coefficients. For field values of energy $E$, these interactions assume the following schematic form,
\be
h ^k \ell^{\,n}
\sim\sum_{s=0}^k \sum_{r=0}^n \frac{\alpha^{s-r}E^{k+n}}{\mpl^{k+n}}
\sim \frac{E^{k+n}}{\mpl^{k+n}}\Big(\alpha^{-n}+ \alpha^{-n+1}+\hdots +\alpha^{k-1}+ \alpha^{k}\Big)
 \,.
\ee
Since here we are only interested in the dependence on $E$, $\alpha$ and $\mpl$, we have dropped all the numerical factors. These do not affect our order of magnitude estimates in any dangerous way, but may, together with the tensor structure, at most serve to make some specific combinations vanish. Including them could therefore potentially only serve to sharpen the general and conservative remarks we make here. We also note that, according to the results of the previous section, there will in general be no terms with $r+s=1$ present since the bimetric action expanded in mass eigenstates contains no terms linear in $\dM_{\mu\nu}$. Nevertheless we will keep these terms for now since they do not influence the general arguments made in this section.

For~$\alpha\ll1$, it is clear that the most suppressed vertices come from pure $\hmn$ terms ($n=0$) and the most enhanced vertices come from pure $\lmn$ terms ($k=0$). At order $m$ in the fluctuations we thus get vertices with the following structure,
\beqn\label{strov}
V_m\sim
\sum_{k=0}^m h ^k \ell^{\,m-k}\sim\left(\frac{E}{\alpha\mpl}\right)^{m}\Big(1+ \alpha+\hdots +\alpha^{2m-1}+ \alpha^{2m}\Big)\,,
\eeqn
where $V_m$ denotes a general vertex with $m$ powers of the field fluctuations appearing. Terms with two derivatives, coming from the Einstein-Hilbert terms, will have a similar structure. Their explicit form turns out not to be relevant for the argument of maintaining perturbativity and we therefore focus on the expansion of the interaction potential (but comment on their inclusion towards the end of this section).

It follows directly from \eqref{strov} that in order for this to define a perturbative expansion, we need to require $E<\alpha\,\mpl$ together with $\alpha<1$. In fact, applying the ($n$th root) Cauchy criterion for convergence show that these requirements are indeed sufficient. Of course, this neglects completely the tensor structure but does provide an estimate for when the expansion is perturbative and the theory is weakly coupled.\\

A subtlety that arises in ordering the expansion properly is that, for small $\alpha$, different orders in fluctuations start to mix with each other at some point, depending on their $\alpha$ dependence. For instance, a fourth order vertex may become of the same perturbative order as a cubic vertex which is multiplied by a higher power of $\alpha$. Therefore, in practice it may be necessary to rearrange the perturbation series differently than in the total number of field fluctuations appearing. This feature is highly dependent on the energy scale under consideration and the exact value of $\alpha$.

 In order to make this more qualitative, let us parameterise the relevant energies as,
\be
E=\alpha^{1+q}\,\mpl\,, 
\qquad
\text{with}~~ q>0\,.
\ee
The condition $q>0$ here simply ensures that we only consider energies which satisfy the condition of perturbativity, $E<\alpha\,\mpl$ for $\alpha<1$. With this parameterisation, for a given $\alpha$, we can probe different energy scales by shifting the value of $q$. We will return to this point further down in this section. In this parameterisation the above vertices read,
\be
V_m\sim
\sum_{k=0}^m  h^k \ell^{\,m-k}
\sim\alpha^{q\,m}  \Big(1+ \alpha+ \alpha^2+\hdots +\alpha^{2m-1}+ \alpha^{2m}\Big)\,.
\ee
In total, the expanded interaction potential therefore consists of vertices $V_m$ given by,
\be\label{defVim}
V_m
\sim V_m^{(0)}+ V_m^{(1)}+ \hdots +V_m^{(2m-1)}+ V_m^{(2m)}\,,
\ee
where  $V_m^{(j)}$ denotes vertices with $m$ powers of fluctuations and $(j+qm)$ powers of~$\alpha$, with $j=0, 1, \dots, 2m$. There is clearly a factor of $\alpha$ suppressing successive terms $V_m^{(j+1)}$ with respect to $V_m^{(j)}$. It is however also clear that a term $V_{m+1}^{(j)}$, with $(m+1)$ fields in the vertex, is suppressed with respect to the term $V_{m}^{(j)}$ with $m$ fields by a factor of $\alpha^q= \frac{E}{\alpha\mpl}$, which, for $q\ll1$, may no longer be small but assume a value rather close to $1$. Depending on the exact value of $q$, the same may be true for $V_{m+2}^{(j)}$, $V_{m+3}^{(j)}$, etc. In particular, at high enough energies, for $q\ll1$, many of these terms end up taking values between $V_{m}^{(j)}$ and $V_{m}^{(j+1)}$. For any given $q\ll1$, we can however always find an integer $p\in\mathbb{N}$ such that $q>1/p$ and the number of such terms is therefore always countable.

In order to elucidate this behaviour more closely for small $q$, i.e.~near the perturbativity bound~$E<\alpha\,\mpl$, let us consider $q=1/p$ with $p\in\mathbb{N}$. It then follows from the definition~\eqref{strov} that the vertices $V_m^{(j)}$ and $V_{m+jp}^{(0)}$ are always of the exact same order. For definiteness, let us also demand that there is at least a factor of $\alpha$ between dominant terms of what we will refer to as different orders. This means that we consider e.g.~$V_{m+k}^{(j)}$, with $k< p$, to be regarded of ``the same order" as $V_{m}^{(j)}$. Note that this is of course somewhat arbitrary since the ratio between neighbouring terms in the rearranged series is typically $\alpha^{1/p}$. Nevertheless, it allows us to rearrange the perturbation series in the following manner:
\begin{align}\label{rearrange}
 & ~~~~V_2^{(0)}+ V_{3}^{(0)}  +\hdots +V_{2+p-1}^{(0)} \nn\\
&+V_2^{(1)} + V_{3}^{(1)}  +\hdots +V_{2+p}^{(1)} +V_{2+p}^{(0)} +V_{2+p+1}^{(0)}+\hdots + V_{2+2p-1}^{(0)}\nn\\
&+V_2^{(2)} + V_{3}^{(2)}  +\hdots +V_{2+p}^{(2)}+V_{2+p+1}^{(1)}+\hdots + V_{2+2p-1}^{(1)}
+V_{2+2p}^{(0)}+\hdots + V_{2+3p-1}^{(0)}\nn\\
&+\hdots\nn\\
&+ V_2^{(j)}+ V_{3}^{(j)}+\hdots +V_{2+p}^{(j)} +V_{2+p+1}^{(j-1)}+\hdots + V_{2+2p-1}^{(j-1)} +\hdots 
+V_{2+jp}^{(0)}+\hdots + V_{2+(j+1)p-1}^{(0)}\nn\\
&+\hdots
\end{align}
Here, every line corresponds to terms of ``the same order" in the perturbative expansion in the following sense: A line with dominant term $V_2^{(j)}\sim\alpha^{j+2/p}$ contains all terms which are of the form $\alpha^{j+(2+k)/p}$ where $k=0,1,2,\dots,p-1$ (note however that for a given line the series is not strictly ordered from left to right since, as noted, e.g.~$V_m^{(j)}$ and $V_{m+jp}^{(0)}$ are always of the exact same order). The next line then starts with dominant term $V_2^{(j+1)}\sim\alpha^{j+1+2/p}$ and so on. Therefore lower lines are always more suppressed than the lines above it. It is important to note that every order contains a finite number of terms. Moreover, from the definition in \eqref{defVim} we have that $V_m^{(j)}=0$ for $j>2m$. Therefore, for example, the 5 first lines are sufficient for considering the influence of higher order terms on the quadratic vertices and the 7 first lines are sufficient for considering the influence on the quadratic and cubic vertices together and so on.

The kinetic terms have been left out of the discussion so far. They contain two derivatives and give, schematically,
\beqn\label{strovd}
\partial^2 h^m&\sim&E^2\left(\frac{E}{\alpha\mpl}\right)^{m}\Big(\alpha^{m}+\alpha^{m+1}+\hdots +\alpha^{2m}\Big)\,,\nn\\
\alpha^2\partial^2 \ell^{\,m}&\sim&E^2\left(\frac{E}{\alpha\mpl}\right)^{m}\Big(\alpha^2 + \alpha^3+\hdots +\alpha^{m+2}\Big)\,.
\eeqn
where we also have taken into account that the $\ell^{\,m}$ vertices coming from the Einstein-Hilbert term for $\fmn$ will have an additional factor of $\alpha^2$ in front. It is clear from the structure of \eqref{strovd} that the terms can be rearranged in the same way as in \eqref{rearrange} to give a valid perturbative expansion for $E<\alpha\,\mpl$. The kinetic and potential terms will indeed have the same general structure in their expansions. In other words, the structure in \eqref{rearrange} represents the expansion of the full bimetric action around maximally symmetric backgrounds in terms of the mass eigenstates.

The above discussion illustrates the behaviour of the expansion near the perturbativity bound $E<\alpha\,\mpl$. From this we can also understand how the higher order vertices start to influence the physics near these energies. In particular, if we required that none of the higher order vertices play any role at the level of cubic interactions, we would have to restrict ourselves to lower energies. The fact that $V_m^{(j)}=0$ for $j>2m$ tell us that the most suppressed cubic term is given by $V_3^{(6)}~\sim\alpha^{6+3q}$. The most dominant quartic term is given by $V_4^{(0)}\sim\alpha^{4q}$. It is therefore enough to demand that $4q>6+3q$, i.e.~$q>6$, to ensure that all the higher order vertices are subdominant to the quadratic and cubic ones (the cubic are then also automatically subdominant to the quadratic terms). In practice, since we wish to probe energies at least up to the mass scale $\mfp=\xi\,\mpl$, this means that if we considered $\xi=\alpha^{1+q}$ with $q>6$, we could be sure that no higher order vertices will influence the physics deduced from studying the full cubic theory.

To illustrate this last point with an example, let us consider the minimal mass for which our spin-2 theory can account for all of the observed DM today (c.f.~section~\ref{sec:pheno}),
\be
\mfp\approx1~\TeV\,,\qquad\Rightarrow\qquad
\xi\approx10^{-15}\,.
\ee
Taking $\xi=\alpha^8$ ($q=7$), this would translate into a bound,
\be
\alpha\gtrsim10^{-15/8}\sim0.01\,.
\ee
With this value for $\alpha$ we could be sure that the cubic theory is enough to discuss physics up to energies $\sim1~\TeV$ and the theory itself remains perturbative up to energies $\sim 0.01\,\mpl$.

As we will see later, our phenomenological analysis reveals that $\alpha$ has to be significantly smaller than 0.01. However, for practical purposes, the perturbativity bound $E<\alpha\,\mpl$ is sufficient to ensure the validity of all our perturbative expressions in section~\ref{sec:pheno}.  This is because the inclusion of (a finite number of) higher-order diagrams merely affects the numerical factors of scattering amplitudes (through additional terms multiplied with positive powers of $\frac{E}{\alpha\,\mpl}<1$). These corrections are irrelevant for our order-of-magnitude estimates. We will comment on this further below.

\subsection{Structure of the cubic vertices in ghost-free bimetric theory}\label{sec:bmcubic}

The detailed expression for the bimetric action~\eqref{bgaction} expanded up to cubic order in the mass eigenstates is provided in Appendix~\ref{app:cubic}. Here we highlight and discuss some of its main characteristics, which also confirms the general arguments given in the previous section.

The dominant structure of the cubic interactions is summarised in Table~\ref{table1}. This table displays the overall coefficients of the various cubic interaction vertices in the $\dG, \dM$ action.

\begin{table}[h!]
\begin{center}
\tabra{1.3}
\begin{tabular}{c|c|c|c}
\toprule
$\delta G^3$~ & ~$\delta G^2 \delta M$~ & $\delta G \delta M^2$ & $\delta M^3$ \\
\hline 
&&&\\
 ~$1, \Lambda$~~ & $0$~ & ~$1,\, \Lambda,\, m_{\rm FP}^2$ ~& ~
 $\alpha,\, \alpha\, \Lambda,\,\alpha\, m_{\rm FP}^2$, 
 $\frac{1}{\alpha},\, \frac{\Lambda}{\alpha},\, \frac{m_{\rm FP}^2}{\alpha}$~\\
 &&&\\
 \bottomrule
\end{tabular}
\end{center}
\caption{Coefficients of cubic interaction vertices (numerical factors neglected) in units of $\mpl^{-1}$. Vertices with a dimensionless coefficient are associated with two derivatives.}
\label{table1}
\end{table}

The first column shows the self-interactions of the massless field. As can be verified from the exact expressions in Appendix~\ref{app:cubicMG} these self-interactions are exactly of standard GR form. This is indeed also true for the quartic vertices, i.e.~the $\dG^4$ terms exactly match the Einstein-Hilbert structure (c.f.~Table~\ref{table2}). This of course implies that $\dG_{\mu\nu}$ gravitates as a massless graviton and is consistent with this being the massless eigenstate since the Einstein-Hilbert structure of GR is fixed uniquely for such a field. It is also consistent with our general arguments of section~\ref{sec:masseigen}.

The second column shows a possible direct decay channel for the massive spin-2 field into two gravitons. Remarkably, these interactions are absent and therefore such a decay is not possible. Again this is consistent with the general arguments of the previous section. Of course, there may still be graviton production due to decay of the massive spin-2 field mediated by SM interactions, but these will generically be heavily suppressed.

The third column displays the gravitational interaction between the massive and massless spin-2 fields. It also captures the tree-level process of the inverse decay discussed later on with respect to possible production mechanisms for the massive spin-2 field. We note that these terms have no $\alpha$ dependence, which already indicates that the massive spin-2 field gravitates with the same strength as SM particles. That the gravitational stress-energy tensor of the spin-2 field indeed coincides with the one obtained via the Noether procedure in flat space follows directly from the general results of ref.~\cite{Leclerc:2005na} which we review in detail in appendix~\ref{app:stressenergy}. For a confirmation of these arguments through an explicit calculation, see \cite{Aoki:2016zgp}.

Finally, the last column displays the self-interactions of the massive spin-2 field. Here we note that there are a variety of terms present and all come with factors of $\alpha$. In particular, in the small $\alpha$ limit, some of these self-interactions will be enhanced as compared to standard GR. This enhancement is particularly strong when $\alpha$ is small and $m_{\rm FP}$ is large.

\begin{table}[h!]
\begin{center}
\tabra{1.3}
\begin{tabular}{c|c|c|c|c}
\toprule
$\dG^4$~ & ~$\dG^3 \dM$~ & $\dG^2 \dM^2$ & $\dG \dM^3$ & $\dM^4$ \\
\hline 
&&&&\\
 ~$1, \Lambda$~~ & $0$~ & ~$1,\, \Lambda,\, m_{\rm FP}^2$ ~& ~
 $\kappa_1,\, \kappa_1\, \Lambda,\,\kappa_1\, m_{\rm FP}^2,$ ~&~
 $\kappa_3,\, \kappa_3\, \Lambda,\,\kappa_3\, m_{\rm FP}^2,$\\
&&&
 $\kappa_2\beta$ ~&~
 $\kappa_4\beta,\,
 \kappa_4\alpha^2m_g^2\beta_2$ \\
 &&&&\\
 \bottomrule
\end{tabular}
\begin{tabular}{c}
$\kappa_1\in\{\alpha^{-1},1,\alpha\}$, $\kappa_2\in\{\alpha^{-3},\alpha^{-1},\alpha\}$, 
$\kappa_3\in\{\alpha^{-2},1,\alpha^2\}$, $\kappa_4\in\{\alpha^{-4},\alpha^{-2},1,\alpha^2\}$\\
\bottomrule
\end{tabular}
\end{center}
\caption{Coefficients of quartic interaction vertices (numerical factors neglected) in units of $\mpl^{-2}$. Vertices with a dimensionless coefficient are associated with two derivatives. In this table we have used the notation $\beta=\alpha^2\mpl^2(\beta_1+\beta_2)/(1+\alpha^2)$ and $m_g^2=\mpl^2/(1+\alpha^2)$.  }
\label{table2}
\end{table}

\section{DM phenomenology}\label{sec:pheno}
We have demonstrated that the heavy spin-2 field of bimetric theory acts as a perfect DM candidate: it interacts extremely weakly with SM particles and gravitates in the same way as ordinary matter does. In addition to these features, for the model to be viable, we need to make sure that the heavy spin-2 field is stable, at least on cosmological time scales, and that its relic abundance can match the one inferred from current observations.

\subsection{Production mechanisms}\label{sec:Mprod}

We first discuss production mechanisms for the heavy spin-2 field abundance possibly active in the early Universe.
Given the extremely weak interactions between the heavy spin-2 field and SM matter, the usual scenario in which the DM relic abundance is built via the freeze-out mechanism cannot be realised since the heavy spin-2 particle is never in thermal equilibrium in the early Universe.  This can be straightforwardly seen by comparing the Hubble rate $H$ to the interaction rate $\Gamma\sim n_\textrm{DM}/\mpl^2$, where $n_\textrm{DM} \sim \rho_\textrm{DM}/\mfp \sim \Omega_\textrm{DM} H^2 \mpl^2/\mfp$ is the DM number density (energy density, density fraction), with the Hubble parameter itself.  The DM abundance is fixed by observations to $\Omega_\textrm{DM}\ll1$ and this means that for $\mfp\geq H$ thermal equilibrium could never be realised. In other words, the Hubble stretch always dominated over the relevant interaction rate $H\gg\Gamma$ and the current DM abundance could not arise via the freeze-out of a thermal DM population.

\paragraph{Gravitational Production.} One possibility is that the heavy spin-2 field could be efficiently produced at the end of inflation, due to the non-adiabaticity of the transition between inflation and the hot Universe. This mechanism is known as gravitational particle production, see~\cite{Chung:1998zb,Kuzmin:1998uv,Chung:2004nh}, and is quite independent of the details of this transition. We will briefly explain why this scenario is not consistent with our setup. 
Gravitational production is most efficient for masses on the order of (or actually, slightly larger than) the Hubble parameter at the end of inflation $H_e$.  The DM abundance today can be written schematically as,
\bea\label{eq:grav_prod}
	10\,\Omega_\textrm{DM} h^2 \approx 10^4 \left(\frac{\mfp}{10^{14}~\GeV}\right)^2\frac{T_\textrm{rh}}{10^7~\GeV}\left(\frac{\mfp}{H_e}\right)^{1/2} e^{-2\mfp/H_e} \, ,
\eea
where $h\sim0.7$ is the little Hubble parameter and $T_\textrm{rh}$ the reheating temperature. Notice however that a large out-of equilibrium component of heavy DM will generate isocurvature perturbations, which are strongly constrained by CMB data~\cite{Ade:2015xua}.  The bound on isocurvature perturbations translates into a lower limit on $\mfp$, or alternatively, on the scale of inflation at thermalisation $H_e$: $\mfp/H_e\gtrsim5$.  This implies that, in order to generate a DM abundance satisfying $\Omega_\textrm{DM} h^2 \sim 0.1$, we find,
\bea\label{eq:gp_mass}
	\mfp\sim 10^{14}\left(\frac{10^7~\GeV}{T_\textrm{rh}}\right)^{1/2}~\GeV \, .
\eea
The lowest mass corresponds to choosing the largest possible reheating temperature, which, for instantaneous reheating, is obtained as $T_\textrm{rh} \sim (H_e\mpl)^{1/2} \leq 10^{16}\, r_s^{1/4}~\GeV$ with $r_s\lesssim0.7$ being the ratio of tensor-to-scalar primordial perturbation amplitudes measured by the Planck experiment~\cite{Ade:2015xua}.  Realistically, the reheating temperature will be at least a factor of a few lower, and the heavy spin-2 mass can be estimated to be at least as heavy as $10^{10}~\GeV$.  As we will show in the following, this requires $\alpha\ll1$, otherwise the spin-2 particle would decay too rapidly. Recall, however, that we cannot take $\alpha$ to be arbitrarily small because our perturbative expansion would otherwise break down as discussed in detail in Sec.~\ref{sec:valexp}.  In fact, we find that this precludes the possibility of gravitational DM production within our framework: As shown in Fig.~\ref{kandinsky}, the region in the $(\alpha,\mfp)$ parameter space where gravitational production is successful is actually excluded by our perturbativity condition.

\paragraph{Freeze-in.} Even if thermal equilibrium is never attained, it is possible to populate the Universe with a nearly decoupled species via a slow ``leakage'' of the thermal bath. This is the so-called freeze-in mechanism~\cite{Hall:2009bx}, which results in a non-thermalised sector, composed of the heavy spin-2 particles in our case.  In this setup two SM particles from the thermal bath annihilate and produce a heavy spin-2 pair via s-channel graviton exchange. Depending on the dynamics of reheating, the generation of DM can proceed either during reheating or in the following radiation dominated era, see~\cite{Garny:2015sjg,Tang:2016vch}. This process is very slow and never counterbalanced by the opposite reaction because the heavy spin-2 abundance remains well below the thermal one at all times.

In our setup, in addition to the usual massless graviton $\dG$ exchange channel, freeze-in can also proceed via exchange of the heavy spin-2 field $\dM$ itself.  The two channels give identical results since the $\alpha$ suppression for the $\textrm{SM}\,\textrm{SM}\rightarrow\dM$ vertex is compensated by the $1/\alpha$ enhancement of the $\dM$ self-interaction $\dM^3$. These production channels are illustrated in Figure~\ref{Fig:treediagrams}.

\begin{figure}
\begin{center}
{\renewcommand{\arraystretch}{0.1}
\begin{tabular}{p{0.45\textwidth} p{0.45\textwidth}}
\begin{center}
\begin{tikzpicture}[thick,
	level/.style={level distance=1.8cm},
	level 2/.style={sibling distance=2.6cm}]
	\coordinate
		child[grow=left]{
			child {
				node {SM}
				edge from parent [sm]
 			}
 			child {
				node {SM}
				edge from parent [sm]
			}
			edge from parent [graviton] node [above=7pt] {$\dG$}
            node [left=50pt] {$\frac{1}{\mpl}$}
            node [right=50pt] {$\frac{1}{\mpl}$}
		}
		child[grow=right, level distance=0pt] {
			child {
				node {$\dM$}
				edge from parent [Mgraviton]
			}
			child {
				node {$\dM$}
				edge from parent [Mgraviton]
			}
		};
\end{tikzpicture}
\end{center}
&
\begin{center}
\begin{tikzpicture}[thick,
	level/.style={level distance=1.8cm},
	level 2/.style={sibling distance=2.6cm}]
	\coordinate
		child[grow=left]{
			child {
				node {SM}
				edge from parent [sm]
 			}
 			child {
				node {SM}
				edge from parent [sm]
			}
			edge from parent [Mgraviton] node [above=7pt] {$\dM$}
            node [left=50pt] {$\frac{\alpha}{\mpl}$}
            node [right=50pt] {$\frac{1}{\alpha\,\mpl}$}
		}
		child[grow=right, level distance=0pt] {
			child {
				node {$\dM$}
				edge from parent [Mgraviton]
			}
			child {
				node {$\dM$}
				edge from parent [Mgraviton]
			}
		};
\end{tikzpicture}
\end{center}
\\
\begin{center}
{\small Freeze-in mediated via $s$-channel exchange of the massless spin-2 $\dG$.}
\end{center}
&
\begin{center}
{\small Freeze-in mediated via $s$-channel exchange of the massive spin-2 $\dM$; the $\alpha$ factors cancel out and the amplitude is the same as for the massless $\dG$ mediation.}
\end{center}
\end{tabular}}
\end{center}
\caption{Tree-level diagrams of s-channel exchanges which are responsible for freeze-in production.}
\label{Fig:treediagrams}
\end{figure}
The generation of DM can be described by a system of coupled Boltzmann equations as in~\cite{Giudice:2000ex}, where the thermally averaged cross section is given by $\la\sigma v\ra\sim T^2/\mpl^4$~\cite{Garny:2015sjg,Tang:2016vch}. The only relevant difference between the two possible production epochs (reheating or radiation domination) is the scaling of the Hubble rate $H\sim \rho^{1/2}$: in the first case $\rho\propto a^{-3/2}$ whereas in the second case $\rho\propto T^2\propto a^{-2}$.  Depending on the efficiency of the reheating process, generally parametrised as $\epsilon_\textrm{rh}^2 = \pi^2 g_* T_\textrm{rh}^4 / 90 \mpl^2 H_e^2 \leq1$ with $g_*=106.75$ being the total number of relativistic degrees of freedom during reheating (which we take to be those of the SM only), the ranges of spin-2 masses for which it is possible to generate the correct DM abundance are,
\bea\label{eq:reh_mass}
	10^4~\GeV \lesssim \mfp \lesssim 10^{17}~\GeV &\qquad& \epsilon_\textrm{rh} =1 \nn\\
	10^7~\GeV \lesssim \mfp \lesssim 10^{16}~\GeV &\qquad& \epsilon_\textrm{rh} =0.1 \\
	10^{10}~\GeV \lesssim \mfp \lesssim 10^{15}~\GeV &\qquad& \epsilon_\textrm{rh} =0.01 \nn\, .
\eea
One can also estimate the total DM abundance directly in radiation domination: matching the observed DM abundance $\Omega_{\rm DM}$ via freeze-in in this case means~\cite{Tang:2016vch},
\begin{equation}\label{eq:freezein_mass}
	m_{\rm FP} \approx \frac{\Omega_{\rm DM} \mpl^3}{\Omega_{\rm b} T_\textrm{rh}^3} m_{\rm p} \eta_{\rm b}\,,
\end{equation}
where $m_{\rm p}$ is the proton mass, $\Omega_{\rm b}$ the abundance of baryons, and $\eta_{\rm b}\approx10^{-9}$ the baryon asymmetry.  Once again, since the scale of inflation cannot be too high in order to avoid overproduction of tensor modes (not observed in the CMB), this implies that the heavy spin-2 mass will be constrained to the range,
\bea\label{eq:fimass}
1~\TeV~\lesssim\mfp\lesssim 10^{11}~\GeV\,.
\eea
This shows that, in principle, freeze-in is a possible production mechanism for our spin-2 DM.
However, we still need to combine this $\alpha$-independent result with the requirement of perturbativity.
As an illustration of the discussion at the end of Sec.~\ref{sec:valexp}, consider the following vertices and their contribution to the production via freeze-in,
\bea
  \textrm{cubic:}& \quad \dG\dM^2: \quad E^3/\mpl\,, \quad &\dM^3: \quad E^3/\alpha\,\mpl\,, \nn\\
  \textrm{quartic:}& \quad \dG\dM^3: \quad E^4/\alpha\,\mpl^2\,, \quad &\dM^4: \quad E^4/\alpha^2\mpl^2\,. \nn
\eea
It is clear that, even for energies below the perturbativity bound, $E<\alpha\,\mpl$, both the $\dM^3$ vertex and the $\dM^4$ vertex dominate over the cubic one, $\dG\dM^2$, for $\alpha<1$.  In order to ensure that the above cubic terms are dominant over the quartic ones, we would have to impose the stronger bound $E<\alpha^2\mpl$.  Were we to go further in the expansion, demanding a ``safe'' order $\alpha$ suppression for everything beyond the cubic order we would recover the condition $E<\alpha^7\mpl$ as derived in section~\ref{sec:valexp}.  
However, when computing the actual production amplitudes, we need to look at the complete diagrams, $\text{SM}\,\text{SM}\rightarrow\dG\rightarrow\dM\,\dM$ and $\text{SM}\,\text{SM}\rightarrow\dM\rightarrow\dM\,\dM$, and compare them to $\text{SM}\,\text{SM}\rightarrow\dM\rightarrow\dM\,\dM\,\dM$.  Then, as we already mentioned, the contribution of the $\dM^3$ vertex to the production rate is the same as the $\dG\dM^2$ one. Moreover, the diagram with the quartic vertex has an additional factor of $\frac{E}{\alpha\,\mpl}$. Hence our perturbative bound $E<\alpha\,\mpl$ is enough to trust our expressions derived with the cubic vertices only.  Higher order vertices will only contribute a finite number of corrections to our estimates, proportional to increasing powers of $\frac{E}{\alpha\,\mpl}<1$, and can thus be safely ignored.

\subsection{Decay and possible signatures}\label{sec:Mdecay}

Since the heavy spin-2 particle does not carry any of the SM charges (which automatically follows from the blindness of gravity to said quantum numbers), it decays universally into all the kinematically allowed channels, i.e.\ into all SM particles $X$ with masses $m_X\leq\mfp/2$. The universality of the decay processes is a feature that our bimetric DM model shares with, for instance, Kaluza-Klein DM~\cite{Han:1998sg}.  However, interestingly, in bimetric DM the massive eigenmode $\delta M_{\mu\nu}$ cannot decay into massless modes.  In other words, there is no graviton production, nor gravitational waves signals, associated to bimetric DM decay.

The decay width into SM particle-antiparticle pairs $X$ is given by~\cite{Han:1998sg},
\begin{align}\label{eq:decayrate}
	\Gamma(\delta M \to XX) = \frac{C_{X}}{80\pi}\frac{\alpha^2\mfp^3}{\mpl^2}
	\,f_X\left(\frac{m^2_X}{\mfp^2}\right) \,,
\end{align}
where the coefficients $C_{X}$ are gathered in Table~\ref{table:cx} and the functions $f_X$ are of the form,
\begin{align}
	f_{V_0} (y) & = 1 \,,\\
	f_{V} (y) & = \left(1 - 4 y\right)^\frac{1}{2}\left(\frac{13}{12} + \frac{14}{39} y + \frac{4}{13} y^2 \right)\,,\\
	f_{f} (y) & = \left(1 - 4 y\right)^\frac{3}{2} \left(1 + \frac 8 3 y\right)\,,\\
	f_{S} (y) & = \left(1 - 4 y\right)^\frac{5}{2}\,,
\end{align}
for massless vector bosons, massive vector bosons, fermions and scalar bosons, respectively.
\begin{table}[htb!]
	
	\label{table:cx}
	\centering
	\begin{tabular}{c|c|c|c|c|c|c|c}
		\toprule
		$\mathbf{X:}$ &
		$\gamma$ &
		$g$ &
		$Z$ &
		$W$ &
		$e$, $\mu$, $\tau$, $\nu_e$, $\nu_\mu$, $\nu_\tau$ &
		$u$, $d$, $c$, $s$, $t$, $b$ &
		$h$
		\\
		\hline
		$\mathbf{C_X:}$ &
		1/2 &
		4 &
		1/2 &
		1 &
		1/4 &
		3/4 &
		1/12 \\
		\bottomrule
	\end{tabular}
	\caption{The coefficients $C_X$ entering eq.~\eqref{eq:decayrate}.}
\end{table}

\begin{figure}[h]
  \centering
    \includegraphics[width=.99\textwidth]{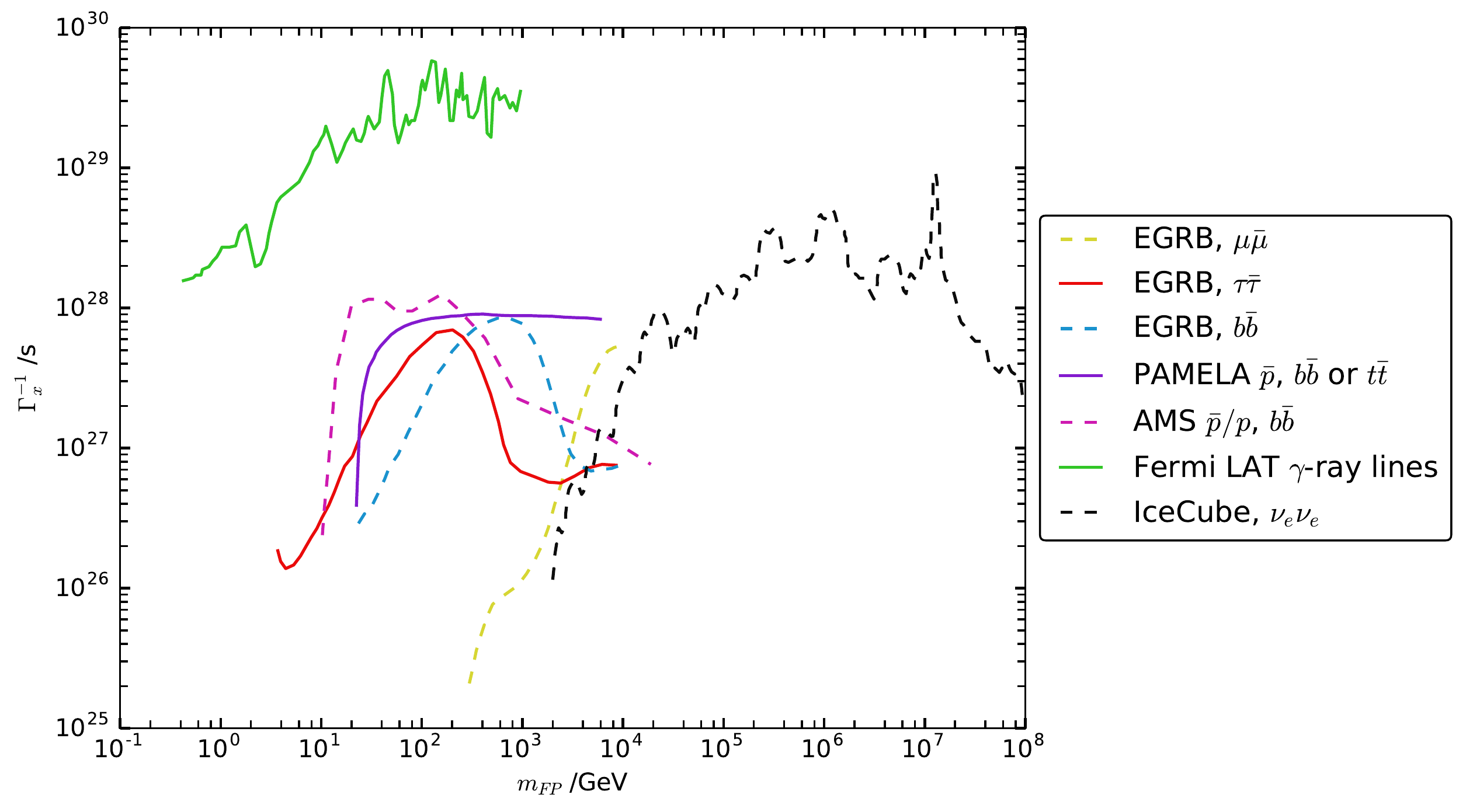}
  \caption{Constraints on the partial decay width from analysis of the EGRB for DM decaying to muons (dashed yellow line), taus (solid red) and $b$ quarks (dashed blue) \cite{Ando:2015qda}. The purple solid line and the violet dashed one show the bounds imposed by the antiproton measurements of the PAMELA \cite{Cirelli:2013hv} and AMS-02 experiments \cite{Giesen:2015ufa}, respectively. We indicate with a solid green line the constraint due to Fermi LAT searches for $\gamma$-ray lines \cite{Ackermann:2015lka} and, with a black dashed line, the bound due to the observation of the high-energy electronic neutrino flux by the IceCube experiment \cite{Aisati:2015vma}.  }
  \label{fig:IDplot}
\end{figure}

The most obvious upper bound on the mass $\mfp$ comes from imposing that the DM be stable on cosmological timescales.  Requiring that its lifetime exceeds the age of the Universe $\tau_U = 13.8$~Gyr implies $\alpha^{2/3} \mfp \lesssim 0.13~\GeV$.  From this constraint we can then derive a consistency upper bound on the DM mass within our perturbative framework. Our expansion (as well as the expression for the width in eq.~\eqref{eq:decayrate}) is valid for $\mfp\leq\alpha\,\mpl$. This limit intersects the bound on the lifetime at $\mfp\approx6.6\times10^6~\GeV$.  Consequently, as remarked before, gravitational particle production is not a viable mechanism to generate the DM as it only operates efficiently for much higher masses.  Furthermore, the viable range for production via freeze-in is shrunk to,
\be\label{eq:mbound}
	1~\TeV \lesssim \mfp \lesssim 6.6\times10^3\TeV\,.
\ee
Given this mass range of the heavy spin-2 field, we can search for distinguishing indirect decay signals. In fact, even tighter constraints than that of eq.~\eqref{eq:mbound} can be derived by using the (non)observation of SM particle fluxes  due to DM decay in different channels.  In general, the constraints on the individual decay widths are heavily dependent on the mass and the propagation properties of the primary and secondary decay products, see for example~\cite{Ibarra:2013cra}.  We gathered the most stringent constraints from DM indirect detection experiments in Fig.~\ref{fig:IDplot}, where we show the bounds on the inverse partial decay widths as a function of the DM mass.

At low DM mass, the strongest constraints for our model come from the Fermi LAT searches for $\gamma$-ray lines \cite{Ackermann:2015lka}: these are the strongest constraints overall, hovering over the $10^{29}$s for the DM lifetime, but only apply up to masses of the order of a TeV.  In the intermediate region, for which $\TeV\lesssim\mfp\lesssim10~\TeV$, the most competitive limits come from the antiproton flux measured by PAMELA \cite{Cirelli:2013hv} instead; the fluxes obtained by the AMS-02 experiment are in the same range \cite{Giesen:2015ufa}.  Moreover, the constraints from the Extragalactic Gamma Ray Background of DM decaying into all SM channels are also in the same ballpark, see \cite{Ando:2015qda} --- we report here only the most significant ones from the muonic, tauonic, and bottom quark channels.  Finally, for the highest mass range we are interested in, $m\gtrsim10~\TeV$, the searches for neutrino lines in IceCube provide the most relevant limits, around $1/\Gamma_{\nu_e}\gtrsim10^{28}$~s \cite{Aisati:2015vma}.

Roughly speaking, we can see that the limit obtained for a DM mass in the range~\eqref{eq:mbound} is approximately 10 orders of magnitude stronger than the bare limit coming from the lifetime of the Universe. In the perturbative regime this translates into an upper limit on the mass of
\be\label{eq:mbound2}
	1~\TeV \lesssim \mfp \lesssim 66~\TeV\,.
\ee
This very limited mass range for heavy spin-2 DM is one of the predictions of our model: a measured DM mass within this narrow range would be a strong indication in support of this model.

\begin{figure}[h]
\begin{center}
\includegraphics[width=1.0\textwidth]{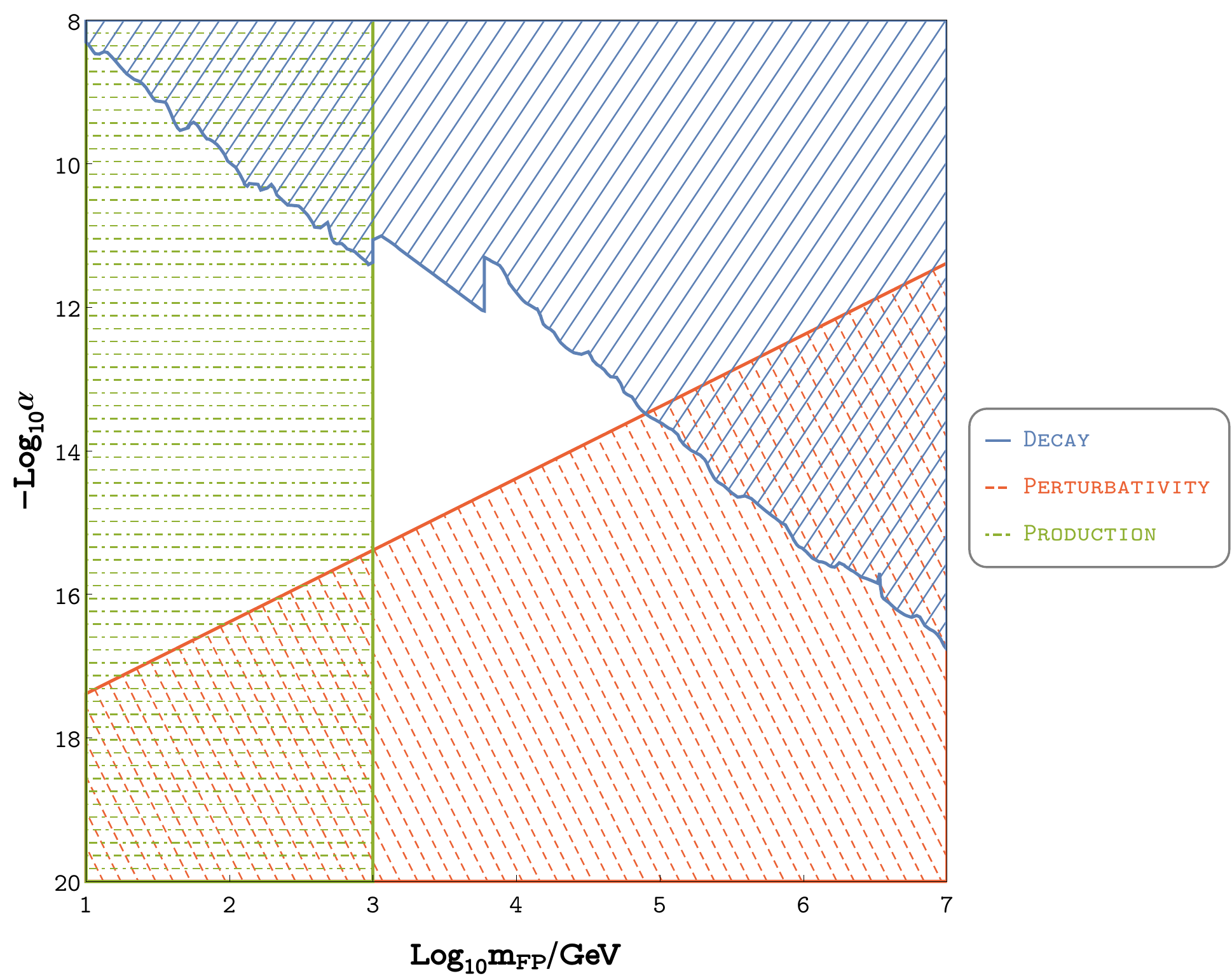}
\end{center}
\caption{Available parameter space for bimetric DM.  The solid blue lines mark the region excluded by the strongest lifetime constraints derived from Fig~\ref{fig:IDplot}; the dashed red lines fill the region we can not study perturbatively; finally, the dot-dashed green lines delineate the range of masses for which the final abundance of DM produced in the early universe can not match the experimentally observed value.}
\label{kandinsky}
\end{figure}

In Fig.~\ref{kandinsky} we collect the strongest constraints on the total decay width, mostly coming from the DM decay into photons and neutrinos, together with the perturbativity limit.  We check them against the different mass ranges available for freeze-in production.  The plot shows the available $(\alpha,\mfp)$ parameter space for bimetric spin-2 DM: the mass of the heavy spin-2 particle is constrained to be in the 1~to 100~TeV range, while the value of $\alpha$ is approximately between $10^{-11}$ and $10^{-15}$. Translating the latter to the value pertaining to the massive spin-2 self-interactions, we find this scale to be in the $10^3~\GeV$ to $10^8~\GeV$ range.

To summarise: The highest and lowest viable values for the mixing parameter $\alpha$ come from intersecting the available mass range for freeze-in production in \eqref{eq:fimass} with the requirement of a long enough DM lifetime and the perturbativity bound, respectively.  The upper limit on the DM mass $\mfp$ is also obtained by joining the latter two constraints: the DM must be stable enough and the energy at production must be within perturbativity.

\section{Discussion}\label{sec:conclusions}

A heavy massive spin-2 field, whose gravitational interactions are described by the ghost-free bimetric theory, possesses all the desired features of a DM candidate. The extremely weak coupling of the spin-2 field to SM matter furthermore explains the absence of DM signals in dedicated detection experiments and collider searches.

The Planck mass $\alpha m_g\sim \alpha\mpl$ of the second metric can range from $1$ to $10^4\,$TeV and the Fierz-Pauli mass for the spin-2 field is on the order of $\mfp\in[1,66]$~TeV. This narrow mass range for the DM candidate is one of the most distinct features of our model. Note however that the upper bound was obtained from the requirement of remaining in the perturbative framework. In principle, non-perturbative methods (which are presently unknown) could reveal that a larger spin-2 mass is also consistent with phenomenology.

Another exceptional property is the universal decay of DM into all SM particles along with the absence of a decay channel into massless gravitons at tree level. Our scenario predicts that mass and interaction scale of the heavy spin-2 field are of the same order of magnitude and only slightly larger than the weak scale. The largeness of the physical Planck mass $\mpl$ is responsible both for suppressing the interactions of DM with baryonic matter and for bringing the theory close to GR. We have therefore not created any new hierarchies of energy scales and moreover related the puzzle of a large Planck scale to the extremely weak interactions of DM.

The above constraints on bimetric parameters are consistent with all current gravity tests. 
They correspond precisely to the overlap of the regions $\mfp^2\gg\Lambda$ for the spin-2 mass and $\alpha\ll1$ for the ratio of Planck masses discussed in Sec.~\ref{sec:pointsol}. Indeed, the Compton wavelength of the spin-2 field is tiny, approximately between $10^{-19}~\text{cm}$ and $10^{-17}~\text{cm}$, resulting in typical Vainshtein radii of $r_V\sim 10^{-10}~\text{cm}$ for the Sun, and $r_V\sim 10^{-24}~\text{cm}$ for millimetre/sub-millimetre tests of the gravitational inverse-square law. These values ensure the validity of the linear approximation for all possible local gravity tests~\cite{Hoyle:2000cv, Yang:2012zzb}. Relative corrections to GR solutions (the ratio of the fifth force to the Newtonian force) involve a factor of $\alpha^2 \exp(-\mfp r)$, which is extremely small in our setup. 
The largest relative deviation from Newton's law, which would be accessible through the sub-millimetre tests, is thus at most $\sim 10^{-22} \exp(-10^{16})$, which is far beyond the reach of any experiment. In a very similar fashion, the smallness of the parameter $\alpha$ ensures that bimetric predictions for cosmology are essentially indistinguishable from the $\Lambda$CDM model and that perturbations remain stable.
Bimetric theory in the above parameter region therefore resembles GR, but with an additional tensor field behaving like cold DM.

The fact that static spherically symmetric solutions in the weak field approximation resemble GR solutions for $\alpha\gg 1$ and $\mfp^2\gg\Lambda$ may suggest that in this parameter range black holes do not have any specific features which would distinguish them from those of GR. For instance, the instability found for spherically symmetric black holes~\cite{Babichev:2013una,Brito:2013wya} --- which is clearly a distinctive feature of bimetric theory, since in GR such black holes are stable --- disappears for large $\mfp$, since the instability range is limited to $\mfp\lesssim r_S^{-1}$. 
On the other hand, the absence of Birkhoff's theorem in bimetric theory allows for the existence of hairy black holes, see~\cite{Volkov:2012wp,Brito:2013xaa} for particular examples and~\cite{Babichev:2015xha} for a recent discussion. It is therefore feasible that hairy black holes are present in our scenario, possibly giving rise to distinct observational features of bimetric theory. It remains to be answered how different these hairy black holes can be from GR black holes, but this question lies beyond the scope of our paper. 

Another distinct property of spin-2 DM are its enhanced self-interaction terms whose form is fixed by the ghost-free structure of the bimetric potential. Self-interacting DM is known to produce observable effects in collisions of galaxy clusters but so far the constraints are not very stringent~\cite{Harvey:2015hha}. Moreover, DM self-interactions would induce distortions in the DM power
spectrum at small scales while leaving the baryonic one untouched.  However, these interactions are mediated by $\dM$ itself, which confines the effectiveness of the corresponding force to risible length scales in astronomical or cosmological terms.  Thus, current constraints on the self-interaction cross section are of little relevance in our case, despite the strength of the $1/\alpha$ enhancement, since in practice the DM particles never feel each other.

Let us finally point out that, in addition to the aforementioned characteristics of our model, the decay of spin-2 DM would in principle exhibit a slightly different spectrum of secondaries (especially neutrinos), compared to the decay of, for instance, a scalar singlet.  Indeed, since the $\delta M_{\mu\nu}$ field couples directly to the energy-momentum tensor of the SM, the two-vector final state (such as $ZZ$ or $W^+W^-$) will be mostly transversal, that is, will carry spin 2.  This is in contrast to DM being a scalar, since in that case only longitudinal states can be produced (at tree level, but other polarisations can appear through higher dimensional operators).  It can also differ from models with a Kaluza-Klein  massive graviton, since in those constructions the transversal-to-longitudinal ratio can be different (depending on the structure of the extra dimensional setup), see for example~\cite{Han:2015cty}. However, this is a very model-dependent statement and we emphasise that in bimetric theory all predictions are fixed by demanding consistency of the theory.  Once the vector bosons decay, the final spectra of secondary neutrinos will bear information about the spin in the guise of peculiar spectral features, see for instance the discussion in~\cite{Garcia-Cely:2016pse}.  These features, however, are very small and hardly within the reach of current experiments.

\vspace{0.5cm}

\acknowledgments
We thank C.~Garcia-Cely, C.~Deffayet, T.~Delahaye, F.~Hassan, J.~Heeck, N.~Khosravi and M.~Volkov for discussions.
This work was supported by the Russian Foundation for Basic Research Grant No. RFBR 15-02-05038 (EB), by the ERC grants IUT23-6, PUTJD110, PUT 1026 (LM, MR, HV), PUT808 (FU) and through the ERDF CoE program (LM, MR, FU, HV), by ERC grant no.~615203 under the FP7 and the Swiss National Science Foundation through the NCCR SwissMAP (ASM) and by the ERC grant no.~307934 under the FP7/2007-2013 (MvS). 
FU acknowledges the Kyiv Astronomical Observatory, BITP Kyiv, and the Odessa State University Observatory for hospitality while this work was completed.
In the process of checking our calculations, we have used the \textit{xTensor} package~\cite{Brizuela:2008ra} developed by J.-M.~Mart\'{\i}n-Garc\'{\i}a for \textit{Mathematica} (\href{http://www.xact.es}{http://www.xact.es}).

\appendix

\section{Spherically symmetric solutions in massive gravity}\label{app:mg}

We consider here linear Fierz-Pauli massive gravity and review how the vDVZ discontinuity manifests itself in this case~\cite{vanDam:1970vg, Zakharov:1970cc}.
It is possible to recover massive gravity solutions from the general bimetric solutions presented in section~\ref{sec:sperical}. 
The parameter limit,
\begin{equation}\label{mglimit}
	\alpha\to \infty, \qquad
	 \mpl\to \infty, \qquad
	  \frac{\mpl}{\alpha} \sim \text{const.},\qquad
	    m_{\rm FP}\sim \text{const.}
\end{equation}
effectively freezes the dynamics of $\fmn$, leaving only five propagating degrees of freedom, corresponding to the polarisations of the massive graviton. 
The limit (\ref{mglimit}) ensures that the Schwarzschild radius $r_S$ stays finite, while the metric functions $\tilde{\lambda}$ and $\tilde{\nu}$ vanish. 
It is also important to stress that in the linear regime, governed by (\ref{bimulin}), the $1/r$ falloff disappears completely, 
leaving only exponentially decaying solution.  This is in contrast to the bimetric case in the $\alpha\rightarrow 0$ limit, for which the $1/r$ tales in the metric functions $\lambda$ and $\nu$ are dominant. As a consequence, massive gravity is not continuously connected to GR.

In the limit~(\ref{mglimit}) the massless graviton $\dG_{\mu\nu}$ 
effectively decouples from the source term $T_{\mu\nu}$,
and the quadratic action~(\ref{bgaction2}) reduces to the well-known 
Fierz-Pauli action for linear massive gravity,
\begin{align}\label{mgaction2}
S^{(2)}=\int\td^4x\sqrt{|\bar g|}\,\biggl[
\Lag^{(2)}_{\rm GR}(\dM)
-\frac{m_{\rm FP}^2}{4}\left(\dM_{\mu\nu}\dM^{\mu\nu}-\dM^2\right)
+\frac{\alpha}{\mpl}\dM_{\mu\nu}T^{\mu\nu}\biggr]\,.
\end{align}
This theory suffers from the vDVZ discontinuity: 
in the limit $m_{\rm FP}\to 0$ the solutions to the equations following from~(\ref{mgaction2}) 
do not correspond to those of linearised GR.
The discontinuity manifests itself in a solution with a point-like source. 
For an ansatz for the metric perturbation of the form (\ref{ds2g}),
the equations of motion derived from~(\ref{mgaction2}) imply,
\begin{equation}\label{mgsource}
	\nu = - \frac{4 C_3 }{3 r}e^{-m_{\rm FP} r}\,, \qquad
	    \lambda = \frac{2 C_3}{3r}(1+m_{\rm FP}r)e^{-m_{\rm FP}r}\, ,
\end{equation}
where $C_3$ is an integration constant, to be fixed by the matching to the source. 
The same result can be derived from (\ref{bimulin}) by applying the limit (\ref{mglimit}).
For $r \gg m^{-1}_{\rm FP}$ the gravitational force is exponentially suppressed by the Yukawa potential. 
At smaller radii, $r\ll m^{-1}_{\rm FP}$, the metric functions exhibit the correct $r^{-1}$ power-law behaviour, 
but the ratio of the metric functions is $|\nu/\lambda| =2$, whereas in GR this ratio is exactly unity. 
This deviation persists for all radii satisfying $r\ll m^{-1}_{\rm FP}$ and thus constitutes a discontinuity of the zero-mass limit.
As a consequence, the Fierz-Pauli theory does not pass the most basic Solar System tests which constrain $|\nu/\lambda|$ to be close to unity.

\section{The cubic action}\label{app:cubic}
Here we provide the explicit form of the bimetric action expanded up to cubic order in fluctuations around the proportional backgrounds. 

\subsection{Metric fluctuations}
We consider fluctuations around the vacuum solution $\bfmn=\bgmn$, in the form,
\be
\gmn=\bgmn+\hmn\,,\qquad \fmn=\bgmn+\lmn\,.
\ee
The metric determinant expanded up to cubic order is given by,
\be
\sqrt{|g|}=\sqrt{|\bar g|}\,\left[1+\frac1{2}[h]+\frac1{8}\left([h]^2-2[h^2]\right)+\frac1{48}\left([h]^3-6[h][h^2]+8[h^3]\right)+\mathcal{O}(h^4)\right]\,,
\ee
where the square brackets around a tensor denote the trace, e.g.~$[h]\equiv \bar g^{\mu\nu}\hmn$ etc. Obviously the analogous expression for $\sqrt{|f|}$ is obtained by the formal replacement $h\rightarrow\ell$. We also write down the expansion of the square root matrix $S=\sqrt{g^{-1}f}$, which on the proportional backgrounds can be obtained from a formal expansion of the form,
\be
\sqrt{1+x}=\sum_{n=0}^{\infty}{1/2\choose n}x^n\,,
\ee
where the binomial coefficient is given by ${1/2\choose n}=(1/2)_{n}/n!$ (with the Pochhammer symbol $(1/2)_{n}$ representing a falling factorial). Written out in full the expansion to cubic order is given by,
\begin{align}
S_{\mu\nu}=&\bgmn +\frac1{2}\left(\lmn-\hmn\right)
+\frac1{8}\left(3h_{\mu\rho}h^\rho_{~\nu}-3h_{\mu\rho}\ell^\rho_{~\nu}
+\ell_{\mu\rho} h^\rho_{~\nu}-\ell_{\mu\rho}\ell^\rho_{~\nu}\right)\nn\\
&+\frac1{16}\bigl(5h_{\mu\rho}h^{\rho\sigma}\ell_{\sigma\nu}
-5h_{\mu\rho}h^{\rho\sigma}h_{\sigma\nu}
-\ell_{\mu\rho} h^{\rho\sigma}h_{\sigma\nu}
+\ell_{\mu\rho} h^{\rho\sigma}\ell_{\sigma\nu}
+h_{\mu\rho}\ell^{\rho\sigma}\ell_{\sigma\nu}\nn\\
&\qquad\quad
-h_{\mu\rho}\ell^{\rho\sigma} h_{\sigma\nu}
+\ell_{\mu\rho}\ell^{\rho\sigma}\ell_{\sigma\nu}
-\ell_{\mu\rho}\ell^{\rho\sigma}h_{\sigma\nu}\bigr)\,.
\end{align}
Recall the bimetric action absent matter sources,
\be
S[g,f]=m_g^2\int\dd^4x\,\biggl[\sqrt{|g|}R(g)+\alpha^2\sqrt{|f|}R(f)
-2\alpha^2m_g^2\sqrt{|g|}\,V(g,f)\biggr]\,.
\ee
After a lengthy exercise in algebra we find that the complete cubic action, modulo boundary terms, can be written as,
\begin{align}\label{cubicact}
S[h,\ell]=\frac{m_{\mathrm{Pl}}^2}{1+\alpha^2}\int\dd^4x\sqrt{|\bar g|}\biggl[&2(1+\alpha^2)\Lambda
+\Lag^{(2)}_{\mathrm{GR}}(h)+\Lag^{(3)}_{\mathrm{GR}}(h)
+\alpha^2\Lag^{(2)}_{\mathrm{GR}}(\ell)+\alpha^2\Lag^{(3)}_{\mathrm{GR}}(\ell)\nn\\
&+\Lag^{(2)}_{\mathrm{Int}}(h,\ell)+\Lag^{(3)}_{\mathrm{Int}}(h,\ell)\biggr]\,.
\end{align}
Here the first term is just a constant piece which is present for constant curvature backgrounds but does not contribute to the equations of motion. The remaining terms in the first line are given by the following GR-like terms, originating from $\sqrt{|g|}(R(g)-2\Lambda)+\alpha^2\sqrt{|f|}(R(f)-2\Lambda)$ with $\Lambda=\alpha^2m_g^2(\beta_0+3\beta_1+3\beta_2+\beta_3)=m_g^2(\beta_4+3\beta_3+3\beta_2+\beta_1)$,
\begin{align}\label{app:Lag2GR}
\Lag^{(2)}_{\mathrm{GR}}(h)=\frac1{4}\biggl[&\nabla_\rho h\nabla^\rho h
-\nabla_\rho \hmn\nabla^\rho h^{\mu\nu}-2\nabla_\rho h\nabla_\mu h^{\mu\rho}
+2\nabla_\rho\hmn\nabla^\nu h^{\mu\rho}\nn\\
&+2\Lambda\left(h_{\mu\nu}h^{\mu\nu}-\frac1{2}h^2\right)\biggr]\,,
\end{align}
and
\begin{align}\label{app:Lag3GR}
\Lag^{(3)}_{\mathrm{GR}}(h)=\,\frac1{4}\biggl[&h^{\mu\nu}\biggl(
\nabla_\mu h_{\rho\sigma}\nabla_\nu h^{\rho\sigma}-\nabla_\mu h\nabla_\nu h
+2\nabla_\nu h\nabla^\rho h_{\mu\rho}+2\nabla_\nu h_{\mu\rho}\nabla^\rho h
-2\nabla_\rho h\nabla^\rho\hmn\nn\\
&\qquad
+2\nabla_\rho\hmn\nabla_\sigma h^{\rho\sigma}
-4\nabla_\nu h_{\rho\sigma}\nabla^\sigma h_\mu^{~\rho}
-2\nabla^\rho h_{\nu\sigma}\nabla^\sigma h_{\mu\rho}
+2\nabla_\sigma h_{\nu\rho}\nabla^\sigma h_\mu^{~\rho}\biggr)\nn\\
&+\frac1{2}h\,\left(\nabla_\rho h\nabla^\rho h
-\nabla_\rho \hmn\nabla^\rho h^{\mu\nu}-2\nabla_\rho h\nabla_\mu h^{\mu\rho}
+2\nabla_\rho\hmn\nabla^\nu h^{\mu\rho}\right)\nn\\
&-\frac{\Lambda}{3}\left(h^3-6h\hmn h^{\mu\nu}+8h^\mu_{~\rho}h^\rho_{~\nu}h^\nu_{~\mu}\right)\biggr]\,,
\end{align}
The second line in (\ref{cubicact}) contains terms from the interactions. First the mass term is,
\be
\Lag^{(2)}_{\mathrm{Int}}(h,\ell)=\frac{\tilde M^2}{4}\left[h^2-\hmn h^{\mu\nu}
+\ell^2-\lmn \ell^{\mu\nu}+2\left(\hmn\ell^{\mu\nu}-h\ell\right)\right]\,,
\ee
where $\tilde M^2\equiv \alpha^2m_g^2\,(\beta_1+2\beta_2+\beta_3)=m_{\mathrm{FP}}^2\,\alpha^2/(1+\alpha^2)$. And finally, the cubic interactions are,
\begin{align}
\Lag^{(3)}_{\mathrm{Int}}(h,\ell)=&\,\frac{\tilde M^2}{24}\biggl[ h^3-6h\hmn h^{\mu\nu}
+5h^\mu_{~\rho}h^\rho_{~\nu}h^\nu_{~\mu}+2\ell^3-9\ell\lmn\ell^{\mu\nu}
+7\ell^\mu_{~\rho}\ell^\rho_{~\nu}\ell^\nu_{~\mu}\nn\\
&\qquad
-3\biggl(h^\mu_{~\rho}\ell^\rho_{~\nu}h^\nu_{~\mu}+3h^\mu_{~\rho}\ell^\rho_{~\nu}\ell^\nu_{~\mu}
-2h\lmn \ell^{\mu\nu}-\ell\hmn h^{\mu\nu}-2\ell\hmn\ell^{\mu\nu}+h\ell^2\biggr)
\biggr]\nn\\
&+\frac{\beta}{24}\biggl[h^3-3h\hmn h^{\mu\nu}
+2h^\mu_{~\rho}h^\rho_{~\nu}h^\nu_{~\mu}-\ell^3+3\ell\lmn\ell^{\mu\nu}
-2\ell^\mu_{~\rho}\ell^\rho_{~\nu}\ell^\nu_{~\mu}\nn\\
&\qquad
-3\biggl(2h^\mu_{~\rho}\ell^\rho_{~\nu}h^\nu_{~\mu}-2h^\mu_{~\rho}\ell^\rho_{~\nu}\ell^\nu_{~\mu}
+h\lmn \ell^{\mu\nu}-\ell\hmn h^{\mu\nu}-2h\hmn\ell^{\mu\nu}+2\ell\hmn\ell^{\mu\nu}\nn\\
&\qquad\qquad
+h^2\ell -h\ell^2\biggr)
\biggr]\,,
\end{align}
where $\beta\equiv \alpha^2m_g^2(\beta_1+\beta_2)$.

\subsection{Mass eigenstates}\label{app:cubicMG}
Replacing $h$ and $\ell$ by the canonically normalised mass eigenstates $\dG$ and $\dM$ using~\eqref{masseigdef}, the cubic action can be written (omitting now the constant part of the action),
\begin{align}
S[\dG,\dM]=\int\dd^4x\sqrt{|g|}\biggl[&
\Lag^{(2)}_{\mathrm{GR}}(\dG)+\frac{1}{m_{\mathrm{Pl}}}\Lag^{(3)}_{\mathrm{GR}}(\dG)
+\Lag^{(2)}_{\mathrm{GR}}(\dM)+\frac{1-\alpha^2}{\alpha\, m_{\mathrm{Pl}}}\Lag^{(3)}_{\mathrm{GR}}(\dM)\nn\\
&+\Lag^{(2)}_{\mathrm{FP}}(\dM)+\frac{1}{m_{\mathrm{Pl}}}\Lag^{(3)}_{\mathrm{GM}}(\dG,\dM)\biggr]\,.
\end{align}
Here the first line contains the same GR like pieces that we have defined in \eqref{app:Lag2GR} and \eqref{app:Lag3GR}. The second line contains the Fierz-Pauli mass term,
\be
\Lag^{(2)}_{\mathrm{FP}}(\dM)=
-\frac{m_{\mathrm{FP}}^2}{4}\left(\dM_{\mu\nu}\dM^{\mu\nu}-\dM^2\right)\,,
\ee
and the remaining cubic interactions (and self-interactions) are given by
\begin{align}
\Lag^{(3)}_{\mathrm{GM}}=&\,
-\frac{m_{\mathrm{FP}}^2(1+\alpha^2)(\beta_1+\beta_2)}{4\alpha\,\mu^2}\,e_3(\dM)\nn\\
&-\frac{m_{\mathrm{FP}}^2}{24\alpha}\biggl[-2[\dM]^3+9[\dM][\dM^2]-7[\dM^3]\nn\\
&\qquad\qquad
+\alpha\left(-3[\dG][\dM]^2+12[\dM][\dG\dM]+3[\dG][\dM^2]-12[\dG\dM^2]\right)\nn\\
&\qquad\qquad
+\alpha^2\left([\dM]^3-6[\dM][\dM^2]+5[\dM^3]\right)\biggr]\nn\\
&-\frac{\Lambda}{4}\biggl[
[\dG][\dM]^2-4[\dM][\dG\dM]-2[\dG][\dM^2]+8[\dG\dM^2]\biggr]\nn\\
&+\frac{1}{4}\biggl[\dG^{\mu\nu}\biggl(
\nabla_\mu \dM_{\rho\sigma}\nabla_\nu \dM^{\rho\sigma}-\nabla_\mu \dM\nabla_\nu \dM
+2\nabla_\nu \dM\nabla_\rho \dM_\mu^{~\rho}+2\nabla_\nu \dM_\mu^{~\rho}\nabla_\rho \dM\nn\\
&\qquad
-2\nabla_\rho \dM\nabla^\rho\dM_{\mu\nu}
+2\nabla_\rho\dM_{\mu\nu}\nabla_\sigma \dM^{\rho\sigma}
-4\nabla_\nu \dM_{\rho\sigma}\nabla^\sigma \dM_\mu^{~\rho}
-2\nabla_\rho \dM_{\nu\sigma}\nabla^\sigma \dM_\mu^{~\rho}\nn\\
&\qquad
+2\nabla_\sigma \dM_{\nu\rho}\nabla^\sigma \dM_\mu^{~\rho}\biggr)\nn\\
&\qquad
+\frac1{2}\dG\,\biggl(\nabla_\rho \dM\nabla^\rho \dM
-\nabla_\rho \dM_{\mu\nu}\nabla^\rho \dM^{\mu\nu}-2\nabla_\rho \dM\nabla_\mu \dM^{\mu\rho}
+2\nabla_\rho\dM_{\mu\nu}\nabla^\nu \dM^{\mu\rho}\biggr)
\biggr]\nn\\
&+\frac{1}{2}\biggl[
\dM^{\mu\nu}\biggl(
\nabla_\mu\dG_{\rho\sigma}\nabla_\nu\dM^{\rho\sigma}
-\nabla_\mu\dG\nabla_\nu\dM
+\nabla^\rho\dG_{\rho\mu}\nabla_\nu\dM
+\nabla_\nu\dG_{\mu\rho}\nabla^\rho\dM\nn\\
&\qquad
-\nabla_\rho\dG_{\mu\nu}\nabla^\rho\dM
+\nabla_\rho\dG^{\rho\sigma}\nabla_\sigma\dM_{\mu\nu}
-2\nabla_\mu\dG^{\rho\sigma}\nabla_\sigma\dM_{\nu\rho}
+\nabla_\mu\dG\nabla^\rho\dM_{\rho\nu}\nn\\
&\qquad
+\nabla^\rho\dG_{\mu\nu}\nabla^\sigma\dM_{\rho\sigma}
-2\nabla_\rho\dG_{\mu\sigma}\nabla_\nu\dM^{\rho\sigma}
-2\nabla^\rho\dG_{\mu\sigma}\nabla^\sigma\dM_{\nu\rho}
+2\nabla^\rho\dG_{\mu\sigma}\nabla_\rho\dM_\nu^{~\sigma}\nn\\
&\qquad
+\nabla^\rho\dG\nabla_\nu\dM_{\mu\rho}
-\nabla^\rho\dG\nabla_\rho\dM_{\mu\nu}
\biggr)\nn\\
&\qquad
+\frac1{2}\dM\biggl(
\nabla_\rho\dG\nabla^\rho\dM-\nabla_\rho\dG_{\mu\nu}\nabla^\rho\dM^{\mu\nu}
-\nabla_\rho\dG\nabla_\sigma\dM^{\rho\sigma}\nn\\
&\qquad
-\nabla_\rho\dG^{\rho\sigma}\nabla_\sigma\dM
+2\nabla_\rho\dG_{\mu\nu}\nabla^\nu\dM^{\mu\rho}
\biggr)
\biggr]\,.
\end{align}
There are no terms of the form $\dG\dG\dM$ present, implying that there is no decay of $\dM$ into massless gravitons at tree level. We also note that there are no $\dG\dG\dG$ terms present and thus all the self-interactions of $\dG$ come from the Einstein-Hilbert term. These are explicit confirmations of the general arguments provided in section~\ref{sec:masseigen}.

\section{Stress energy tensors}\label{app:stressenergy}
Here we provide a general argument for the flat space on-shell {\it physical equivalence} between Noether and gravitational stress energy tensors, following~\cite{Leclerc:2005na}.
Consider a gravitational Lagrangian of the following general form,
\be
\Lag_{\rm Tot} = \Lag_{\rm GR}(G)+\Lag_{\rm m}(G,\partial G, \dM,\nabla\dM)\,.
\ee
As our notation suggests, the total Lagrangian $\Lag_{\rm Tot}$ consist of a gravitational part $\Lag_{\rm GR}$ for the nonlinear metric $G$ which has the form of the standard Einstein-Hilbert term and a ``matter" part $\Lag_{\rm m}$. The matter field $\dM$ couples to the metric $G$ as well as its first derivatives (which must enter in the form of Christoffel symbols due to covariance). 

As we discussed in section~\ref{sec:masseigen}, the above Lagrangian captures the structure of bimetric theory expanded to infinite order around proportional backgrounds in terms of the linear massive mode $\dM$ and the nonlinear gravitational metric $G_{\mu\nu}=\bar{g}_{\mu\nu}+\dG_{\mu\nu}/\mpl$. The general results derived in the following therefore all hold true in our setup.

We compute the gravitational stress energy tensor corresponding to the above theory from,
\be\label{app:Tgrav}
T^{\mu\nu}=-\frac{1}{\sqrt{|G|}}\frac{\delta(\sqrt{|G|}\Lag_{\rm m})}{\delta(G_{\mu\nu})}
=-\frac{1}{\sqrt{|G|}}\left[\frac{\p(\sqrt{|G|}\Lag_{\rm m})}{\p G}
-\p_\rho\left(\frac{\p(\sqrt{|G|}\Lag_{\rm m})}{\p(\p_\rho G_{\mu\nu})}\right)\right]\,.
\ee
Note that the last term is usually absent since standard matter only couples to the graviton and not its derivatives. 

On the other hand, the Noether stress-energy tensor derived from translational symmetry in flat space is,
\be\label{app:Tnoether}
\tau^{\rho}_{\ph{\rho}\nu}=\frac{\left.\p\Lag_{\rm m}\right|_{G=\eta}}{\p(\p_\rho\dM_{\mu\sigma})}\p_\nu\dM_{\mu\sigma}
-\delta^\rho_\nu\left.\Lag_{\rm m}\right|_{G=\eta}\,.
\ee
As a final ingredient we recall the Euler-Lagrange equations for the matter field in the form,
\be\label{app:ELeqs}
\p_\rho\left(\frac{\p(\sqrt{|G|}\Lag_{\rm m})}{\p(\p_\rho\dM_{\mu\nu})}\right)=\frac{\p(\sqrt{|G|}\Lag_{\rm m})}{\p\dM_{\mu\nu}}\,.
\ee
Now we evaluate the gradient,
\begin{align}
\p_\rho(\sqrt{G}\Lag_{\rm m})&=
\frac{\p(\sqrt{G}\Lag_{\rm m})}{\p\dM_{\mu\nu}}\p_\rho\dM_{\mu\nu}
+\frac{\p(\sqrt{G}\Lag_{\rm m})}{\p(\p_\sigma\dM_{\mu\nu})}\p_\rho\p_\sigma\dM_{\mu\nu}\nn\\
&\quad+\frac{\p(\sqrt{G}\Lag_{\rm m})}{\p G_{\mu\nu}}\p_\rho G_{\mu\nu}
+\frac{\p(\sqrt{G}\Lag_{\rm m})}{\p(\p_\sigma G_{\mu\nu})}\p_\rho\p_\sigma G_{\mu\nu}\,.
\end{align}
Using the definition~\eqref{app:Tgrav} and the equations of motion~\eqref{app:ELeqs} we can use this to obtain,
\be
\p_{\rho}\left[\sqrt{|G|}\tau^\rho_{\ph\rho\nu}\right]
-\sqrt{|G|}T^{\rho\sigma}\p_\nu G_{\rho\sigma}
+\p_\rho\left(\frac{\p(\sqrt{|G|}\Lag_{\rm m})}{\p(\p_\rho G_{\mu\sigma})}
\p_\nu G_{\mu\sigma}\right)
=0\,.
\ee
Further manipulation and using the covariant conversation law $\nabla_\rho T^\rho_{\ph\rho\nu}=0$ finally gives,
\be\label{app:TTrel}
\p_{\rho}\left[\sqrt{|G|}\left(\tau^\rho_{\ph\rho\nu}-2T^\rho_{\ph\rho\nu}\right)
+\frac{\p(\sqrt{|G|}\Lag_{\rm m})}{\p(\p_\rho G_{\mu\sigma})}\p_\nu G_{\mu\sigma}\right]
=0\,.
\ee
This is the general correspondence between the stress energy tensors in curved spacetime for a theory of the form we are considering. Note that in general the stress energy tensors are of course not equal and furthermore, the Noether stress energy tensor is strictly speaking derived from flat space considerations. In order to check the physical implications of~\eqref{app:TTrel} it is instructive to study the momenta defined by,
\be
\pi_\nu=\int\td^3x\sqrt{|G|}\,\tau^0_{\ph0\nu}\,,\qquad
P_\nu=2\int\td^3x\sqrt{|G|}\,T^0_{\ph0\nu}\,.
\ee
We see that~\eqref{app:TTrel} implies that (the constant of integration can be shown to vanish),
\be
\pi_\nu=P_\nu-\int\td^3x\sqrt{|G|}\,
\frac{\p(\sqrt{|G|}\,\Lag_{\rm m})}{\p(\p_0 G_{\mu\sigma})}\p_\nu G_{\mu\sigma}\,.
\ee
Therefore, in flat space where $\p_\nu G_{\mu\sigma}=0$, we find that $\pi_\nu=P_\nu$ which shows the physical equivalence between the prescriptions. From \eqref{app:TTrel} it also follows that in flat space (and/or for theories where matter does not couple to the derivative of the graviton) we always have (again the constant of integration can be shown to vanish),
\be
\int\td^4x\left(\tau^\rho_{\ph\rho\nu}-2T^\rho_{\ph\rho\nu}\right)=0\,.
\ee
Hence, in flat space the integrated stress energy tensors are always equal on-shell (modulo a factor of 2 which is due to our definition of $T_{\mu\nu}$). 

This last relation has an immediate and interesting implication. Namely, considering a non-relativistic gas (fluid) of matter satisfying Bose statistics, the Noether stress energy is known to correspond to a stress energy of the form of dust $\tau^\rho_{\ph\rho\nu}\sim\mathrm{diag}(\rho,0,0,0)$. The above relations now imply that this is also the form of the source for the gravitational field and therefore the massive spin-2 field will behave just as cold DM non-relativistically. This can also be explicitly verified directly from our expressions for the cubic interaction terms in appendix~\ref{app:cubic}.\footnote{We note that in Ref.~\cite{Aoki:2016zgp} an explicit calculation of this within an analogous setup was carried out which confirms this. As our general arguments show, this must always be true for any theory of the form we are considering.}


\end{document}